\documentclass[preprint,showpacs,preprintnumbers,amssymb,superscriptaddress,aps,prd,nofootinbib,11pt]{revtex4-1}
\usepackage{graphicx, epsfig, amssymb} 
\usepackage{amsmath, amsfonts}
\usepackage{bm} 
\usepackage{enumitem}
\usepackage[usenames]{color}
\definecolor{navyblue}{rgb}{0.0, 0.0, 0.5}
\usepackage[linktocpage,colorlinks=true,allcolors=navyblue]{hyperref}
\usepackage[caption=false]{subfig}
\usepackage[utf8]{inputenc}
\usepackage{tensor}
\usepackage{soul}
\usepackage{pict2e}
\usepackage{diagbox}

\usepackage{float}
\usepackage{comment}
\usepackage{array}
\usepackage{units}
\usepackage{wasysym}
\def\p{\partial}

\newcommand{\nn}{\nonumber}
\newcommand{\Thetageo}{\Theta_{\text{geo}}}

\newcommand{\bbow}{b_r}
\newcommand{\thetabow}{\theta_r}

\renewcommand{\vec}[1]{\mathbf{#1}}

\newcommand{\Ft}{\tilde{F}_{lm}}
\newcommand{\St}{\tilde{S}_{lm}}
\newcommand{\Ftc}{\tilde{F}}
\newcommand{\Stc}{\tilde{S}}
\newcommand{\Qt}{\tilde{Q}_{lm}}
\newcommand{\w}{\omega_0}

\begin{document}\title{\large Rainbow scattering of gravitational plane waves by a compact body}

\author{Tom Stratton}\email{tstratton1@sheffield.ac.uk}
\affiliation{Consortium for Fundamental Physics,
School of Mathematics and Statistics,
University of Sheffield, Hicks Building, Hounsfield Road, Sheffield S3 7RH, United Kingdom}

\author{Sam R. Dolan}\email{s.dolan@sheffield.ac.uk}
\affiliation{Consortium for Fundamental Physics,
School of Mathematics and Statistics,
University of Sheffield, Hicks Building, Hounsfield Road, Sheffield S3 7RH, United Kingdom}

\begin{abstract} 
We study the time-independent scattering of a planar gravitational wave propagating in the curved spacetime of a compact body with a polytropic equation of state. We begin by considering the geometric-optics limit, in which the gravitational wave propagates along null geodesics of the spacetime; we show that a wavefront passing through a neutron star of tenuity $R/M = 6$ will be focussed at a cusp caustic near the star's surface. Next, using the linearized Einstein Field Equations on a spherically-symmetric spacetime, we construct the metric perturbations in the odd and even parity sectors; and, with partial-wave methods, we numerically compute the gravitational scattering cross section from helicity-conserving and helicity-reversing amplitudes. At long wavelengths, the cross section is insensitive to stellar structure and, in the limit $M \omega \rightarrow 0$, it reduces to the known low-frequency approximation of the black hole case. At higher frequencies $M \omega \gtrsim 1$, the gravitational wave probes the internal structure of the body. In essence, we find that the gravitational wave cross section is similar to that for a massless scalar field, although with subtle effects arising from the non-zero helicity-reversing amplitude, and the coupling in the even-parity sector between the gravitational wave and the fluid of the body. The cross section exhibits \emph{rainbow scattering} with an Airy-type oscillation superposed on a Rutherford cross section. We show that the rainbow angle, which arises from a stationary point in the geodesic deflection function, depends on the polytropic index. In principle, rainbow scattering provides a diagnostic of the equation of state of the compact body; but, in practice, this requires a high-frequency astrophysical source of gravitational waves.
\end{abstract}

\date{\today}

\maketitle

\section{Introduction}

The first direct detection of gravitational waves (GW) from a binary black hole (BH) inspiral was announced in 2016 \cite{Abbott:2016blz}, and the first catalogue of Gravitational Wave Transients (GWTC-1) was released in late 2018 \cite{LIGOScientific:2018mvr}. The catalogue comprises ten confirmed transient events, of which nine are consistent with the GW signal generated by a binary black hole merger \cite{Abbott:2016nmj,Abbott:2017vtc,Abbott:2017gyy,Abbott:2017oio}; and one (GW170817) is consistent with that generated by a binary neutron star inspiral \cite{TheLIGOScientific:2017qsa}. The latter was accompanied by a series of observations across the electromagnetic spectrum \cite{GBM:2017lvd}.

Gravitational waves (GWs) are generated by highly energetic astrophysical processes, such as binary mergers and supernovae. GWs are subject to negligible scattering from the intervening dust, gas and plasma between the source and Earth, due to their weak coupling to the matter sector. GWs provide observers with relatively direct access to the physics at the heart of the source; in contrast, electromagnetic signals are much more strongly affected by intervening matter. On the other hand, by the equivalence principle, GWs feel the gravitational influence of matter/energy sources, and so they can be significantly scattered in strongly-curved regions of spacetime, such as near black holes or neutron stars.

A gravitational wave scattered by a neutron star bears the imprint of its gravitational potential. Consequently, observations of scattered waves could, in principle, probe and inform models of the Equation of State (EoS) of nuclear matter under immense pressures. In practice, such scattered waves would be challenging to observe. 

This study aims to improve our theoretical understanding of gravitational wave scattering in a idealised scenario, in which a monochromatic gravitational wave of circular frequency $\omega$ impinges upon a spherically-symmetric compact body of radius $R$ and mass $M$ in vacuum. We model the compact body using three different polytropes, with indices $n \in \{0,0.5,1\}$ (see below). We focus particularly on computing the scattering cross section $d \sigma / d \Omega$, i.e., the intensity of the flux scattered to infinity as a function of scattering angle. 


The time-independent scattering of waves by a black hole has received attention since 1968, following the work of Matzner \cite{Matzner:1968} and, later, Sanchez \cite{Sanchez:1977vz}. The work in Refs.~\cite{Handler:1980un,Zhang:1984vt,Matzner:1985rjn} culminated in a 1988 monograph by Futterman, Handler and Matzner \cite{Futterman:1988ni}. In recent years, black hole scattering calculations have been made by Crispino and coworkers \cite{Crispino:2009ki, Crispino:2009xt,  Crispino:2014eea, Crispino:2015gua, Leite:2017zyb} and several other groups \cite{Dolan:2007ut, Dolan:2008kf, Kanai:2013rga, Nambu:2015aea, Rosa:2016bli, Alexandre:2018crg, Sporea:2018rsk, Leite:2019uql}. The idealised scenario of time-independent scattering by compact objects has also received some recent attention \cite{Dolan:2017rtj, Cotaescu:2018etx} (see also related work~\cite{Tominaga:1999iy,Tominaga:2000cs,Bernuzzi:2008rq}). 

The scattering process depends, in part, on the compactness of the scattering body, described here by the (dimensionless) \emph{tenuity} $R c^2/ (GM)$ (henceforth we use units such that $G = c = 1$) and the index $n$ of the polytropic equation of state, with $n=0$ corresponding to a star of constant density. 
 Some characteristic values include $R/M \sim 6$ for neutron stars, $\sim 1.4 \times 10^3$ for a massive white dwarf (e.g.~Sirius B), $\sim 9.4\times10^3$ for a typical white dwarf, $4.7 \times 10^5$ for the Sun, and $1.4 \times 10^9$ for Earth. 
 
\begin{figure}
\centering
  \includegraphics[width=0.6\textwidth]{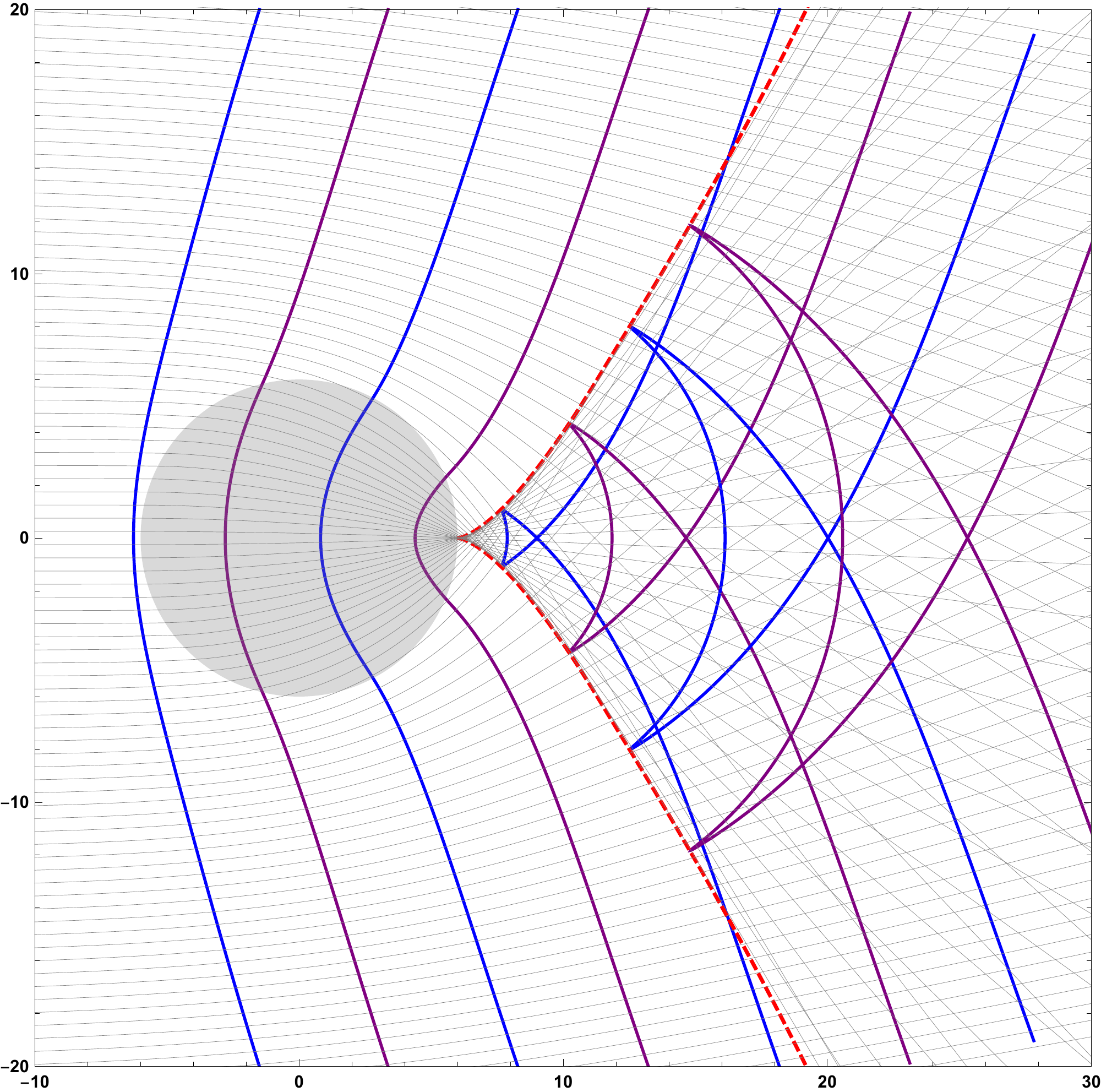}
\caption{Formation of a cusp caustic on a neutron star spacetime. A congruence of null geodesics passing through a compact body are shown as thin grey lines; the successive wavefronts, propagating from left to right, are shown as alternating blue and purple solid lines; and the cusp caustic appears as a dotted red line.}
\label{fig:cuspcaustic}
\end{figure}

\begin{figure}
\centering
\subfloat[]{\label{subfig:n0}%
  \includegraphics[width=0.3\textwidth]{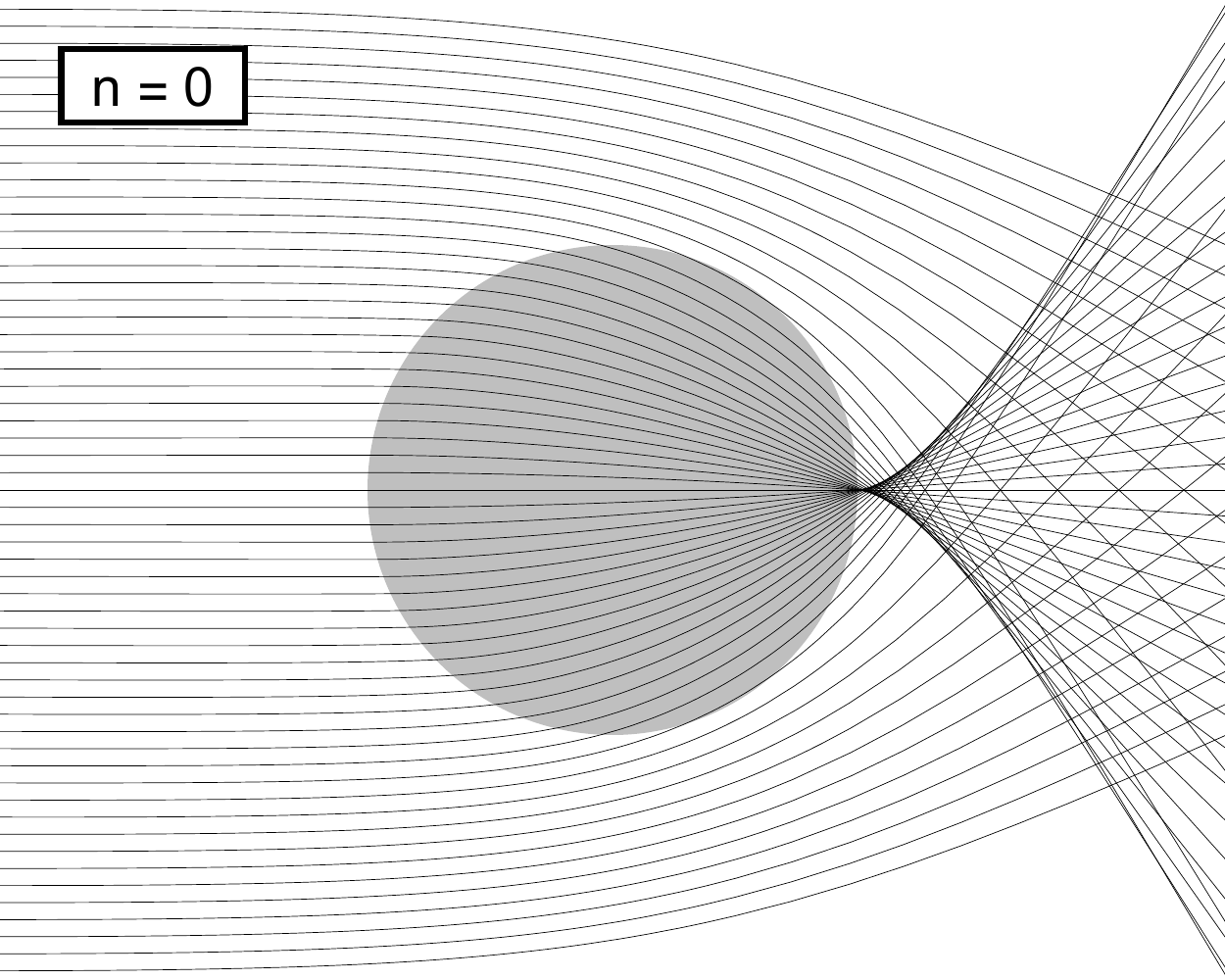}
} 
\hspace{5mm}
\subfloat[]{\label{subfig:npt5}%
  \includegraphics[width=0.3\textwidth]{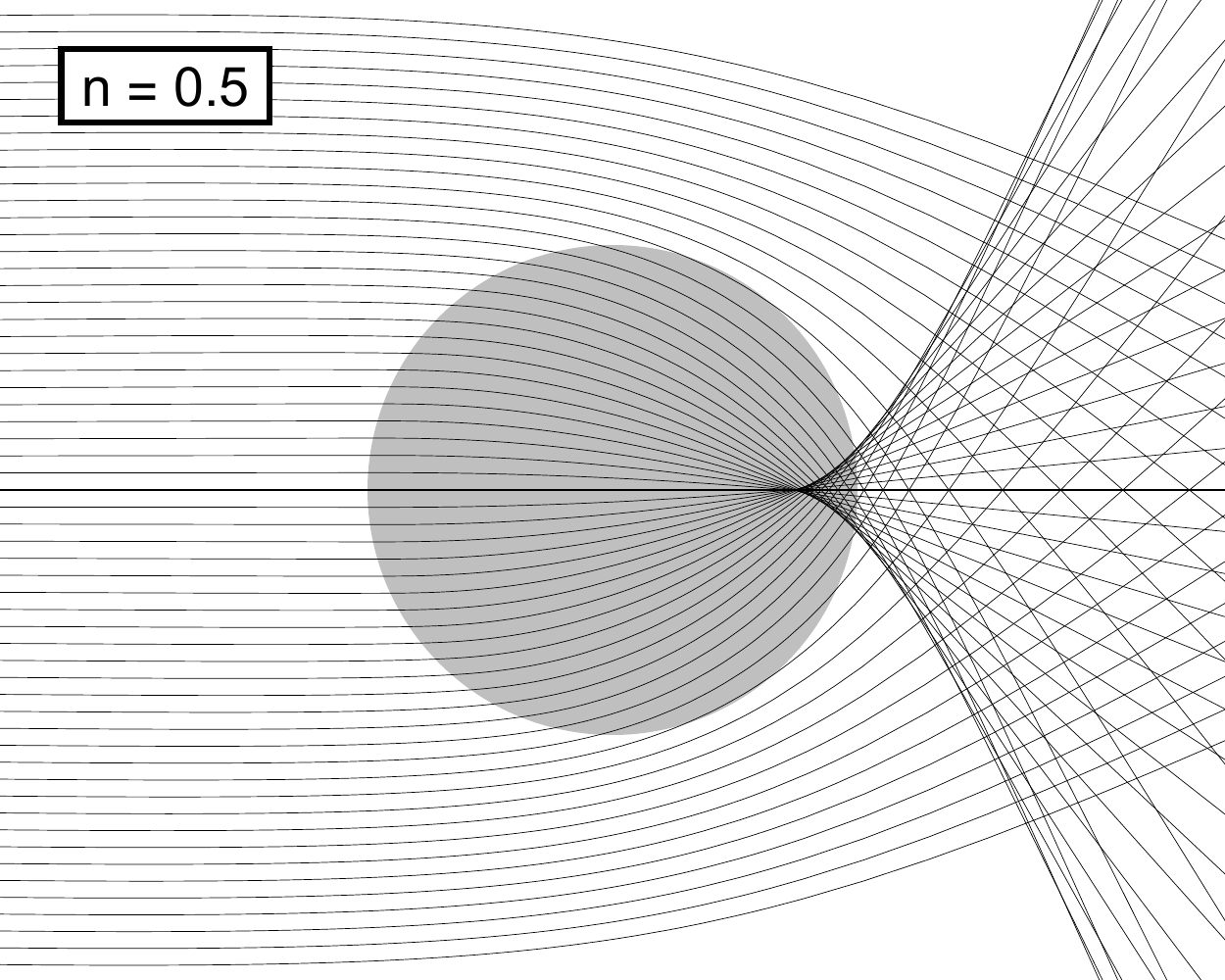}
}
\hspace{5mm}
\subfloat[]{\label{subfig:n1}%
  \includegraphics[width=0.3\textwidth]{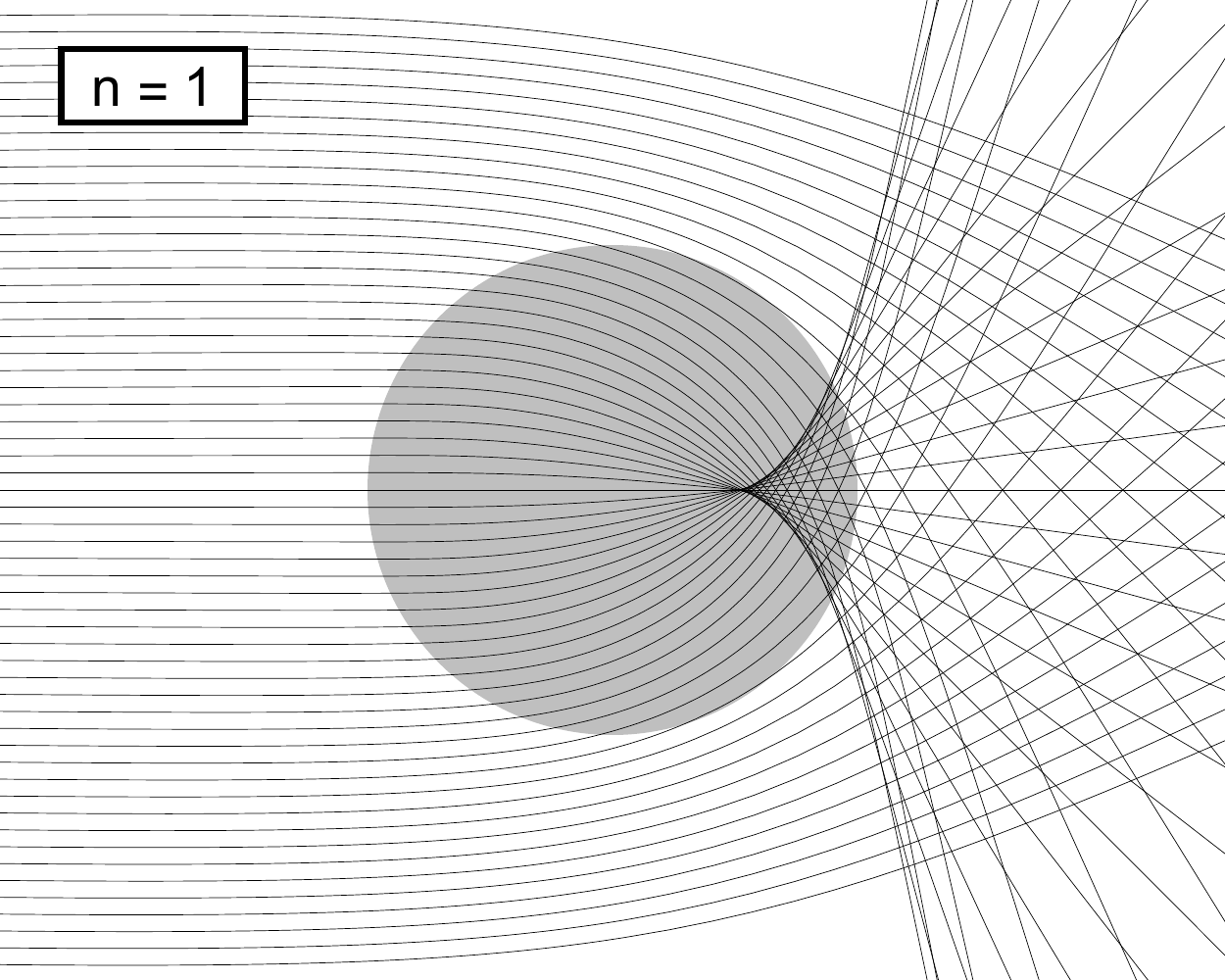}
} \\
\caption{A congruence of null geodesics passing through a spherical compact body of tenuity $R/M = 6$ with a polytropic equation of state index $n=0$ (a), $n=0.5$ (b), and $n=1$ (c). The cusp caustic forms near the surface of the body for $n=0$, and deeper inside the body for higher polytropic indices.  Asymptotically, the cusp caustic defines a rainbow wedge with rainbow angle (a) $\theta_r=59.6^{\circ}$, (b) $\theta_r=66.6^{\circ}$, (c) $\theta_r=79.7^{\circ}$.}
\label{fig:intro_nf}
\end{figure}

\begin{figure}
\centering

\subfloat[]{\label{subfig:wave:npt5}%
  \includegraphics[width=0.45\textwidth]{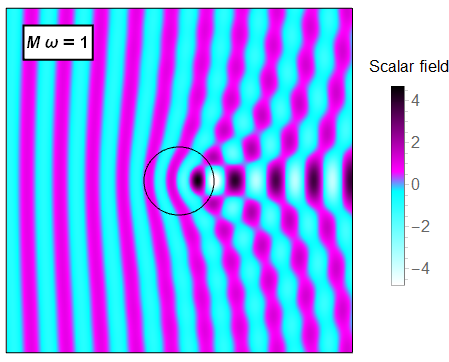}
}
\hspace{10mm}
\subfloat[]{\label{subfig:wave:n1}%
  \includegraphics[width=0.45\textwidth]{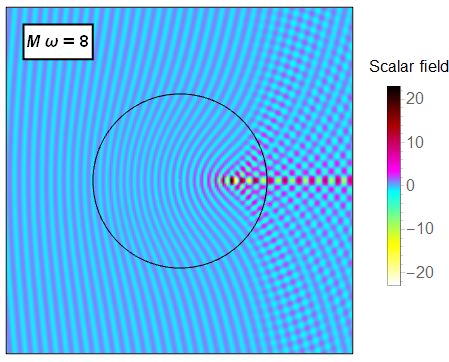}
} \\
\caption{A unit amplitude plane wave consisting of a scalar field being scattered by a compact body (outline in black). The compact body is a polytrope with index $n=1$ and $R/M=6$. The incident wave has coupling (a) $M\omega=1$ and (b) $M\omega=8$. The amplitude can be increased by a factor up to approximately 4 and 20 for $M\omega=1$ and $M\omega=8$ respectively. For the higher frequency case, it is just possible to make out the rainbow scattering feature of a primary peak at $\theta \approx 79.7^{\circ}$.}
\label{fig:NF_scalarMw8}
\end{figure}

In Ref.~\cite{Dolan:2017rtj} (henceforth Paper I) we studied the scattering of a \emph{scalar field} $\Phi$, governed by the Klein-Gordon equation $\Box_g \Phi = 0$, by a spherically-symmetric star of constant density ($n=0$) and mass $M$. We found that, for moderate-to-high frequencies $M \omega \gtrsim 1$, the scattering process may be understood via semi-classical arguments, with reference to a congruence of null geodesics which pass through the star (see Figs.~\ref{fig:cuspcaustic}--\ref{fig:NF_scalarMw8}). In Paper I, we conjectured that gravitational waves would, in essence, behave in a qualitatively similar fashion to massless scalar waves, with additional features relating to spin transport, helicity-reversal and coupling to matter degrees of freedom. We investigate this conjecture here, building on the foundation laid in the works of Ipser and Price \cite{Ipser:1991ind}; Kojima \cite{Kojima:1992ie}; Allen \emph{et al.} \cite{Allen:1997xj}; Martel \& Poisson \cite{Martel:2005ir}; Barack \& Lousto \cite{Barack:2005nr}; and others \cite{Cunningham:1978zfa,Cunningham:1979px,Chandrasekhar247,Chandrasekhar247b,Chandrasekhar449}. 

Key features of the scattering process, in the semiclassical picture, are summarized in Figs.~\ref{fig:cuspcaustic}, \ref{fig:intro_nf} and \ref{fig:NF_scalarMw8}. Figure \ref{fig:cuspcaustic} shows a congruence of null geodesics, initially parallel, encountering a constant density star of tenuity $R/M = 6$. The rays come together at a cusp caustic, which may be inside or outside the neutron star. Beyond the cusp, each wavefront has multiple segments. An observer on-axis downstream would encounter the wavefront in two parts; arriving first, the segment that scattered from the weak-field potential, and later, due to time dilation, the segment that passed through the central potential. Figure \ref{fig:intro_nf} shows that the position of the cusp caustic, and the \emph{rainbow angle} $\theta_r$ of the wedge, depends on the polytropic index $n$ of the matter distribution. The caustic moves closer to the centre, and the wedge gets wider, as the body becomes more centrally dense (see also Fig.~1 in Paper I and Fig.~2 in Ref.~\cite{Halder:2019cmp} for higher tenuities).  Figure \ref{fig:NF_scalarMw8} shows the wave scattering pattern for a scalar field $\Phi$ at moderate ($M\omega = 1$) and high frequencies ($M \omega = 8$). In the latter case, significant amplification at the cusp caustic is evident. 

Key features of the scattering cross section $d\sigma / d\Omega$ may be understood with reference to the deflection function $\Thetageo(b)$, where $b$ is the impact parameter of a ray (a null geodesic). The `classical' scattering cross section, 
$
\left. \frac{d \sigma}{d \Omega} \right|_{\text{cl}} = \frac{b}{\sin \theta \, \left| \Theta'  \right|} 
\label{eq:classical}
$
with $\Theta' = \frac{d \Theta}{d b}$, 
is singular at the poles ($\theta = 0$, $\pi$), and also at stationary points of the deflection function, if they exist. In semiclassical theory \cite{Airy:1838,Ford:1959}, the singularities soften into familiar interference effects: glories arise near the poles, and stationary points in the deflection function generate rainbow scattering oscillations. 
The standard semiclassical prescription for rainbow scattering (reviewed in Paper I) leads to Airy's formula \cite{Airy:1838},
\begin{equation}
\frac{d\sigma}{d\Omega} \approx \frac{2 \pi \bbow}{\omega q^{2/3} \sin \theta} \text{Ai}^2\left( \frac{ \theta - \thetabow }{q^{1/3}} \right), \quad \quad q \equiv \frac{\Theta''_r}{2 \omega^2} ,  \label{eq:airy}
\end{equation}
where the condition $\Theta^\prime(\bbow) = 0$ defines a rainbow impact parameter $\bbow$, a rainbow angle $\thetabow \equiv |\Theta(\bbow)|$, and a second derivative $\Theta_r'' \equiv \tfrac{d^2\Theta}{db^2}(\bbow)$. Thus, the colours of the rainbow are separated in angle according to wavelength, with the `primary' peak appearing at $ \thetabow - 0.237 [\lambda^2 \Theta_r'']^{1/3}$ (see Eq.~(\ref{eq:airy})), where $\lambda$ is the wavelength. The scattered intensity falls off rapidly in the classically-forbidden shadow region for $\theta > \theta_r$, whereas on the bright side of the rainbow the cross section has supernumerary peaks beyond the primary. 

At low frequencies $M\omega \ll 1$, we anticipate that the scattering cross section for massless fields encountering a compact body will be the same as for fields approaching a Schwarzschild black hole, namely \cite{Westervelt:1971pm, Peters:1976jx, DeLogi:1977dp, Dolan:2007ut},
\begin{align}
\lim_{M\omega \rightarrow 0} \left(M^{-2} \frac{d\sigma}{d\Omega} \right) = \begin{cases} \frac{\cos^{4s}(\theta / 2)}{\sin^4(\theta/2)}  \quad  & s = 0, \, \frac{1}{2} \; \text{and} \; 1,  \\[5pt] 
\frac{ \cos^8(\theta/2) + \sin^8(\theta/2) }{\sin^4(\theta/2)} \quad & s = 2 \, , \label{eq:lowfreq_grav_scat} 
\end{cases}
\end{align}
The first line states the general result for spin 0 (scalar), $1/2$ (spinor), and $1$ (electromagnetic) waves. The second line is the result for a gravitational wave ($s=2$). The extra `anomalous' term arises from a scattering amplitude associated with the reversal of the helicity of the incident wave (see e.g.~\cite{Dolan:2007ut}).

This paper is organised as follows. In Sec.~\ref{sec:methods} we describe our methods, focussing on the stellar model \ref{subsec:model}; the gravitational perturbations outside the star \ref{subsec:exterior}, inside the star \ref{subsec:interior} and at the surface \ref{sec:perts_surface}; the construction of a physical solution \ref{subsec:solution}; the gravitational plane wave \ref{subsec:planewaves}; the scattering cross section \ref{subsec:crosssection}; and the numerical methods employed \ref{subsec:nummethod}. In Sec.~\ref{sec:results} we present a selection of results. In Sec.~\ref{sec:conclusions} we conclude with a discussion of physical implications, analogies, and future work. The $-+++$ signature and units such that $G = c = 1$ are used throughout. Commas and semi-colons are used to denote  partial and covariant derivatives, respectively.

\section{Model and Methods\label{sec:methods}}

We model time-independent scattering of a gravitational plane wave by a spherically symmetric compact body, such as a neutron star. Standard perturbation theory is employed, where the metric is expressed as $g_{\mu\nu}^T = g_{\mu\nu} + h_{\mu\nu}$. The background metric, $g_{\mu\nu}$, is a known solution to the Einstein field equations (EFE). The gravitational perturbation, $h_{\mu\nu}$, is governed by the linearised perturbed EFE. Many authors have studied perturbations of the Schwarzschild (exterior) spacetime \cite{Regge:1957td,Zerilli:1971wd,vishveshwara1970scattering}. Typically a spherical harmonic decomposition is used, and possibly a Fourier decomposition. We employ a powerful gauge-invariant and covariant formalism for metric perturbations on Schwarzschild, developed by Martel and Poisson \cite{Martel:2005ir}.

The primary object of interest is the scattering cross section $d\sigma / d\Omega$: the flux of the scattered radiation per unit solid angle. To calculate it, we follow the standard approach in \cite{Futterman:1988ni}: we compute $h_{\mu\nu}$ and `match' it to a plane wave in the far field, extracting the scattered part of the wave.  The details of our calculation differ from the standard approach which employs Weyl scalars and the Newman-Penrose formalism \cite{Futterman:1988ni}. Instead, we work with the metric perturbation directly, using the Martel-Poisson formalism. 

\subsection{Metric and stellar model\label{subsec:model}}

The line element for the background space-time of a spherically-symmetric star in Schwarzschild coordinates $\{t,r,\theta,\phi\}$ may be written as
\begin{equation}
ds^2 = g_{\mu \nu} dx^\mu dx^\nu =  g_{ab} dx^a dx^b + r^2 \Omega_{AB} dx^A dx^B.
\end{equation}
Here, lower case Latin indices run over $\{t,r\}$, upper case Latin indices run over $\{\theta,\phi\}$, and Greek indices run over all coordinates (unless specified otherwise). We have
\begin{equation}
g_{ab} dx^a dx^b = -A(r) \, dt^2 + B^{-1}(r) \, dr^2, \quad \quad \Omega_{AB} dx^A dx^A =  d\theta^2 + \sin^2 \theta \,d\phi^2,
\end{equation}
where $A(r)$ and $B(r)$ are radial functions that depend on the matter distribution. 
Lower and upper case Latin indices are lowered with $g_{ab}$ and $\Omega_{AB}$ respectively, and raised with the corresponding inverse metrics. Greek indices are lowered and raised with $g_{\mu\nu}$ and its inverse $g^{\mu \nu}$.

A spherically-symmetric star composed of an ideal fluid has the stress energy tensor
\begin{equation}\label{stress-energy-perfect-fluid}
T_{\mu\nu} = (\rho + p) u_\mu u_\nu + p g_{\mu \nu}
\end{equation}
where $u^\mu(r)$ is the fluid 4-velocity, $p(r)$ the pressure, and $\rho(r)$ the energy density. It is convenient to define a function $m(r)$ in terms of the metric function $B$ via
\begin{equation} \label{eqn:def_mass_func}
B(r) = 1 - \frac{2m(r)}{r}.
\end{equation}
The $(t,t)$ and $(r,r)$ components of the EFE give 
\begin{equation}\label{eqn:TOVconstraint}
\frac{1}{A} \frac{d A}{dr} = -\frac{2}{ \rho + p} \frac{dp}{dr},
\quad \quad
\frac{dm}{dr} = 4 \pi r^2 \rho .
\end{equation}
From the conservation of energy-momentum, $\tensor{T}{^{\mu\nu}_{;\nu}}=0$, one can then derive the Tolman-Oppenheimer-Volkov (TOV) equation of hydrostatic equilibrium for the interior of the star,
\begin{equation}\label{eqn:TOV}
\frac{dp}{dr} = - \frac{(\rho + p)(m + 4\pi r^3 p)}{r(r-2m)}.
\end{equation}
For details see e.g.~Chap.~10 of Schutz \cite{Schutz:1985jx}.

Outside the star, the mass is constant and thus $A(r)=B(r)=1-2M/r$, by Birkhoff's theorem. Inside the star, we must specify an Equation of State (EoS) before we can solve Eqs.~(\ref{eqn:TOVconstraint})-(\ref{eqn:TOV}) to find $p(\rho)$. 
 A simple and effective model is a polytropic star, with EoS
\begin{equation}
p(\rho) = \kappa \rho^{1+1/n},
\end{equation}
where $n$ is the polytropic index and $\kappa$ is a constant. 
By solving (\ref{eqn:TOV}) numerically for a given $n$, $\kappa$ and central density $\rho(0)$, we obtain the pressure and density profiles, $p(r)$ and $\rho(r)$, and also the radius and mass of the star, $R$ and $M$. 

The speed of sound in the star is
\begin{equation}\label{eqn:speedofsound}
c_s^2 = c^2 \frac{ \partial p}{ \partial \rho} \bigg|_S,
\end{equation}
where the derivative is taken at constant entropy, $S$.  For a typical white dwarf or neutron star the temperature is effectively zero everywhere, so that the specific entropy is also negligible everywhere \cite{Shapiro:1983du}. 

As $n \rightarrow 0$ the fluid that makes up the star becomes stiffer. The case $n=0$ corresponds to a star of constant density  (Schwarzschild's interior solution \cite{Schwarzschild:1916}) with an infinite speed of sound \cite{1964Tooper}. Keeping in mind its paradoxical nature, the constant density star is nevertheless interesting as a limit of the sequence of decreasing $n$. In the $n=0$ case, the solution can be found analytically. The metric functions for a constant density star are
\begin{eqnarray}
A(r) &=& \frac{1}{4R^3} \left( \sqrt{R^3 - 2Mr^2} - 3 R \sqrt{R - 2M} \right)^2 , \nn \\
B(r) &=& 1 - \frac{2Mr^2}{R^3} .
\end{eqnarray}
For the polytropes with $ n \neq 0$, we found it necessary to match the numerical solution to an analytical polytropic `atmosphere' \cite{Ipser:1991ind}. To obtain more suitable neutron star models, we took $n \in \{0.5,1\}$ (see for example \cite{Allen:1997xj}). Higher values of $n$ are typically used to model more diffuse stars.

\subsection{The exterior perturbation\label{subsec:exterior}}

We now consider the metric perturbation equations in three regions: in the vacuum exterior; in the interior of the compact body; and at its surface. It is important to note that the metric perturbation itself, denoted $h_{\mu \nu}$, is not gauge invariant. Under a small coordinate transformation $x^{\mu} \rightarrow x'^{\mu} = x^{\mu} + \xi^{\mu}$, where $\xi^{\mu} = \mathcal{O}({\epsilon})$ is small, the linear metric perturbation transforms according to $h_{\mu\nu} \rightarrow h_{\mu\nu}' = h_{\mu\nu} - 2  \xi_{(\nu;\mu)}$. Different gauges have different technical benefits. For example, the asymptotic plane wave solution (Sec.~\ref{subsec:planewaves}) is expressed in transverse-traceless gauge, whereas it is simplest to solve for the metric perturbation in the stars interior in Regge-Wheeler (RW) gauge (Sec.~\ref{subsec:interior}). The main quantity of interest in this work -- the energy flux of the scattered radiation -- is gauge-invariant. 


Far from the compact body in asymptotically Cartesian coordinates, as $r \rightarrow \infty$, the radiative parts of the metric perturbation in spherical coordinates scale as \cite{Martel:2005ir}
\begin{equation}
h_{ab}^{\text{rad}} \sim r^{-1}, \quad h_{aB}^{\text{rad}} \sim r^{0}, \quad h_{AB}^{\text{rad}} \sim r^1. \label{eq:radiative_scalings}
\end{equation}
Working in a specific radiation gauge in which $t^a h_{ab} = 0 = t^a h_{a B}$, where $t^a$ is the timelike Killing vector, Martel and Poisson \cite{Martel:2005ir} showed that the radiative part of a metric perturbation on Schwarzschild can be written as a sum over modes, as
\begin{equation}\label{eqn:total_rad_pert}
h_{AB}^{\text{rad}} = r \sum_{p=\pm 1} \sum_{l \geq2}^{\infty} \sum_{m=-l}^{m=l}   \Phi_{lm}^{p}(r,t) X_{AB}^{lmp}(\theta,\phi) , 
\end{equation}
where $X_{AB}^{lmp}$ are the tensor spherical harmonics, and $\Phi_{lm}^{\pm}$ are the master functions defined in \cite{Martel:2005ir}\footnote{Martel and Poisson denote $X_{AB}^{lm+}$ and $X_{AB}^{lm-}$ by $X_{AB}^{lm}$ and $Y_{AB}^{lm}$ respectively}. They also show that the leading order parts of $h_{ab}^{\text{rad}}$ and $h_{aB}^{\text{rad}}$ are zero in this gauge. The parity of a mode is either polar/even ($p=1$) or axial/odd ($p=-1$). The polar and axial master functions, $\Phi_{lm}^{+}$ and $\Phi_{lm}^{-}$, are also known as the Zerilli-Moncrief and Cunningham-Price-Moncrief master functions, respectively \cite{Martel:2005ir}. These functions may be decomposed in Fourier modes $\tilde{\Phi}_{lm}^{\pm}(r,\omega)$ using
\begin{equation}
\Phi_{lm}^{\pm}(r,t) = \frac{1}{2\pi}\int_{-\infty}^{\infty} d\omega \, \tilde{\Phi}_{lm}^{\pm}(r,\omega) e^{i \omega t} .
\end{equation}

The master functions $\tilde{\Phi}_{lm}^{\pm}$, which do not depend on the choice of gauge, are governed by a pair of second order ODEs, namely
\begin{equation} \label{eq:ODE_exterior}
\frac{d^2 \tilde{\Phi}_{lm}^{\pm}}{d r_*^2} + \left( \omega^2 - V_{l}^{\pm}(r) \right) \tilde{\Phi}_{lm}^{\pm} = 0.
\end{equation}
Here $V_l^-(r)$ is the Regge-Wheeler potential,  
\begin{equation}
V_l^-(r) := A \left( \frac{l(l+1)}{r^2} - \frac{6M}{r^3} \right),
\end{equation}
and $V_l^+(r)$ the Zerilli potential 
\begin{equation}
V_l^+(r) := \frac{A}{\Lambda^2}  \left[  \mu^2  \left(\frac{\mu+2}{r^2}+\frac{6
   M}{r^3}\right) + \frac{36 M^2}  {r^4} \left(\mu +\frac{2 M}{r}\right) \right],
\end{equation}
with $\mu :=(l-1)(l+2)$ and $\Lambda :=\mu + 6M/r$. The tortoise coordinate $r_*$ is defined by 
\begin{equation}
\frac{dr}{dr_*} = \sqrt{AB}.
\end{equation}

In the far field, the solutions to Eq.~(\ref{eq:ODE_exterior}) behave asymptotically as
\begin{equation}\label{eq:bc_farfield}
\tilde{\Phi}_{lm}^{p}(r,\omega) \sim A_{lp}^{\text{in}}(\omega) e^{i \omega r_*} + A_{lp}^{\text{out}}(\omega) e^{-i \omega r_*},
\end{equation}
where $ A_{lp}^{\text{in}}$ and $ A_{lp}^{\text{out}}$ are complex constants.
The full radiative perturbation in the far field can be reconstructed as
\begin{equation}\label{eq:AinAout}
h_{AB}^{\text{total}}  \sim r \int_{-\infty}^{\infty} d \omega \left[ \sum_{p=\pm 1} \sum_{l \geq2}^{\infty} \sum_{m=-l}^{m=l}   \left( A_{lp}^{in}(\omega) e^{i \omega r_*} + A_{lp}^{out}(\omega) e^{-i \omega r_*} \right) e^{i \omega t} X_{AB}^{lmp} \right].
\end{equation}

\subsection{The interior perturbation\label{subsec:interior}}

Although there are works on gauge-invariant formalisms for a general spherically symmetric spacetime \cite{Seidel:1987in,Seidel:1990xb}, we found it convenient to work in Regge-Wheeler (RW) gauge. In particular, we make use of the formalisms of Kojima \cite{Kojima:1992ie}, and Allen \emph{et al.}~\cite{Allen:1997xj}, for the odd and even parity sectors respectively. The even (odd) parity sector couples (does not couple) to fluid perturbations \cite{thorne1967non}.

\subsubsection{Odd parity}

In RW gauge, the odd-parity perturbations for a general spherically symmetric spacetime are determined by a single scalar function, $\Qt(r,\omega)$, used in the decomposition of $h_{\mu\nu}$ (see for example \cite{Kojima:1992ie} where $\Qt \rightarrow X_{lm}$, or \cite{thorne1967non}). This function is governed by the radial equation 
%
%
%
\begin{equation}\label{eq:ODE_odd_interior}
\frac{d^2 \Qt}{d r_*^2} + \left( \omega^2 - \mathcal{V}_{l}(r) \right) \Qt = 0, 
\end{equation}
where
\begin{equation}
\mathcal{V}_l(r) := A(r) \left[ \frac{l(l+1)}{r^2} - \frac{6m(r)}{r} - 4 \pi (p_{\text{rad}} - \rho)  \right] .
\end{equation}
%
%
In vacuum, $\mathcal{V}_l(r)$ reduces to $V_l^-$, and thus $\Qt$ satisfies the Regge-Wheeler equation and it is proportional to $\tilde{\Phi}_{lm}^-$, 
\begin{equation}\label{eq:Q_CPM}
\tilde{\Phi}_{lm}^{-}(r,\omega) =  \frac{2}{i \omega} \Qt.
\end{equation}
\subsubsection{Even parity}
The even-parity spacetime perturbations couple to the fluid perturbations. Ipser and Price reduced the problem to a fourth-order system of ODEs working in frequency space \cite{Ipser:1991ind}, or alternatively a coupled pair of second-order equations. Allen \emph{et al.}~investigated even-parity perturbations as an initial value problem \cite{Allen:1997xj}, also in RW gauge. They formulated the problem as three second-order wave equations and a constraint equation. Two of these equations are for space-time variables, $S_{lm}(r,t)$ and $F_{lm}(r,t)$; see Eqs.~(10) and (11) of \cite{Allen:1997xj} for their relations to the metric perturbation. The constraint can be used to eliminate the third variable,  describing the fluid perturbations, again reducing the problem to two coupled wave equations (Eqs.~(14) and (18) of \cite{Allen:1997xj}). Here we make use of these two coupled equations. The Fourier transforms of the variables of Allen \emph{et al.}, $\St(r,\omega)$ and $\Ft(r,\omega)$, are governed by 
%
\begin{multline}
\frac{d^2 \Stc}{ d r_*^2} + \left[ \omega^2 + \frac{A}{r^3}  \left( 4 \pi r^3 \left( \rho + 3 p \right) + 2 m - l(l+1)r \right) \right]    \Stc  \\ 
=  -\frac{4 A^2 }{r^5} \left[ \frac{ \left( m + 4 \pi p r^3 \right)^2 }{ \left( r – 2m \right) } + 4 \pi \rho r^3 - 3m \right] \Ftc, \label{eqn:interiorS}
\end{multline}
and
%
%
\begin{multline}
\frac{d^2 \Ftc}{ d r_*^2}  - \left( 1 - \frac{1}{c_s^2} \right)  \sqrt{\frac{A}{B} } \frac{1}{r^2}  \left( m + 4\pi p r^3 \right)  \frac{d \Ftc}{ d r_*}  \\ 
+ \left[ \frac{\omega^2}{c_s^2} + \frac{A}{r^3}  \left( 4 \pi r^3 \left( 3 \rho + \frac{p}{c_s^2} \right) – m \left( 1 - \frac{3}{c_s^2} \right)  - l(l+1)r \right) \right]   \Ftc  \\ 
 = \left( 1 - \frac{1}{c_s^2} \right) r \sqrt{\frac{B}{A}} \frac{d \Stc}{ d r_*} + \left[ 2B + \left( 1 - \frac{1}{c_s^2} \right) \frac{l(l+1)}{2}  - 8\pi( p + \rho) r^2 \right] \Stc, \label{eqn:interiorF}
\end{multline}
where we have dropped mode labels $lm$ for brevity, and here $m=m(r)$ refers to the mass function defined in  Eq.~(\ref{eqn:def_mass_func}). 

In vacuum, the even-parity master function $\tilde{\Phi}_{lm}^{+}$ is related to $\St$ and $\Ft$ by
\begin{equation}\label{eq:SF_ZM}
\tilde{\Phi}_{lm}^{+}(r,\omega) = \frac{2}{l(l+1)} \left[ \Ft + \frac{2}{\Lambda  } \left( 2 A \Ft - rA \p_r \Ft + r^2 \St \right)  \right].
\end{equation}

\subsection{Perturbations at the stellar surface} \label{sec:perts_surface}

The continuity of the first and second fundamental forms at the stellar surface $r=R$ implies the continuity of $\Qt,\St,\Ft,\p_r \Qt,$ and $\p_r \St$ there, and a junction condition for $\Ft$ \cite{Allen:1997xj,Ipser:1991ind},
\begin{multline} \label{eq:JCdF}
[\p_r \Ft]^+_-  =  - \frac{\rho}{2 \omega^2 r^2 (p+\rho) } \bigg[  r \left( l(l+1) - 2 \omega^2 r^2  A^{-1} \right) \St  + r^2 l(l+1) \p_r \St  \\
  +  \left( \left(  l(l+1) - \omega^2 r^2 A^{-1} \right) \p_r A - 2 \omega^2 r^2 \right) \Ft + 2 \omega^2 r^2 \p_r \Ft  \bigg] \Bigg|_{r=R^-}.
\end{multline}
%
%
The notation $[z(r)]^+_-$ denotes $\text{lim}_{\epsilon \rightarrow 0} [z(R + \epsilon )-z(R - \epsilon)]$ and the right-hand side of Eq.~(\ref{eq:JCdF}) is evaluated on the surface by taking the limit from below. 

In addition, on the stellar surface the Lagrangian change in pressure should vanish \cite{Ipser:1991ind}. This yields a boundary condition in the form of a single constraint relating $\St$, $\p_r \St$, $\Ft$ and $\p_r \Ft$ just inside the stellar surface (see Eq.~(5.2) in \cite{Ipser:1991ind}). This constraint is equivalent to the radial derivative of the Hamiltonian constraint in \cite{Allen:1997xj}. 

\subsection{Construction of a physical solution\label{subsec:solution}}
In the odd-parity sector, we start with a solution that is regular at the origin, $\Qt \sim r^{l+1}$, and use the differential equation (\ref{eq:ODE_odd_interior}) to extend it into the exterior, using the continuity of $\Qt$ and its derivative across the stellar surface. Outside the star, we may use Eq.~(\ref{eq:Q_CPM}) to relate $\Qt$ to the master variable $\tilde{\Phi}_{lm}^{-}$.

In the even-parity sector, there are two independent solutions to Eqs.~(\ref{eqn:interiorF}) and (\ref{eqn:interiorS}) that satisfy regularity at the origin,
%
\begin{align}
\{ \tilde{S}_{lm}^1, \tilde{F}_{lm}^1 \}  &\sim \{ r^{l+1}, a_1 r^{l+3} \}, \\ 
\{ \tilde{S}_{lm}^2, \tilde{F}_{lm}^2 \}  &\sim \{ r^{l+3}, a_2 r^{l+1} \} \quad \text{as} \quad r \rightarrow 0,
\end{align}
where $a_{1}$ and $a_{2}$ are constants determined by the metric/stellar model. We may write our solution as a linear sum
$
\mathbf{Y} = \alpha_1 \mathbf{Y}_1 + \alpha_2 \mathbf{Y}_2 , 
$ 
where $\vec{Y}= [\St,\p_r \St, \Ft, \p_r \Ft ]^t$, and $\vec{Y}_i$ refer to the two independent solutions above. Just inside the stellar surface, we must apply the boundary condition that arises from insisting that the Lagrangian change in pressure vanishes (Eq.~(5.2) in \cite{Ipser:1991ind}). This constraint yields a unique $\Ft$ and $\St$ that satisfies the boundary conditions at the stellar surface and the origin, up to an overall scaling. We may then extend this solution beyond the surface using the junction conditions of Sec.~\ref{sec:perts_surface}.

\subsection{Plane waves} \label{subsec:planewaves}

A left circularly-polarized gravitational plane wave of angular frequency $\w$ travelling up the $z$ axis and expressed in a spherical coordinate system, $\{t,r,\theta,\phi \}$, on flat space in transverse-traceless gauge is 
\begin{equation}\label{eqn:h_components_planewave}
\renewcommand*{\arraystretch}{1.5}
h_{ij} = \text{Re} \left\{ H e^{i( 2\phi - \chi)} \begin{bmatrix}
				
				 s^2     & r s c    & i r s^2  \\
				 \bullet & r^2 c^2  & i r^2 sc \\
				 \bullet & \bullet &  -r^2 s^2  \\
				
				\end{bmatrix} \right\}.
\end{equation}
%
%
where $H$ is the amplitude of the wave; $\chi \equiv \w (t-z)$; indices $(i,j)$ run over spatial coordinates; $s$ and $c$ are shorthand for $\sin \theta$ and $\cos \theta$, respectively. All other components are zero. 

The physical solution we seek is, qualitatively, the sum of a plane wave and an outgoing radiative component in the far field. However, a plane wave is not a valid solution on the Schwarzschild background, even in the far-field region, due to the long-range $1/r$ nature of the field. Following convention (see \cite{Futterman:1988ni}) we replace (\ref{eqn:h_components_planewave}) with a distorted plane wave by making the substitution $r \rightarrow r_*$ in the exponent of Eq.~(\ref{eqn:h_components_planewave}) (i.e. $z \rightarrow z_* = r_* \cos \theta$). From this metric perturbation, we may compute the master variables $\Phi^{\text{plane}}_{l m p}$. 

We construct our solution for $\Phi_{lmp}$ so that its ingoing part matches, asymptotically, the ingoing part of the distorted plane wave $\Phi^{\text{plane}}_{lmp}$. The outgoing, scattered component of the radiation is then 
\begin{equation}\label{eq:asymp_matching_masters}
\Phi^{\text{scat}}_{lmp} = \Phi_{lmp} - \Phi^{\text{plane}}_{lmp}. 
\end{equation} 
The master functions for a left-handed circularly-polarized distorted plane wave can be expanded in the far field as
\begin{eqnarray}
\Phi^{\text{plane}}_{l2,-1}(r,t) & = & \frac{2 \pi  H C_{l2} }{\w} \left( (-1)^{l+1} e^{-i \w r_*} +  e^{i \w r_*} \right)  e^{-i\w t} + \mathcal{O}\left(r^{-1} \right)  , \label{eqn:odd_master_function_PW} \\
\Phi^{\text{plane}}_{l2,+1}(r,t)  & = & -\frac{2 \pi i H C_{l2} }{\w} \left((-1)^{l+1} e^{-i \w r_*} +  e^{i \w r_*} \right)  e^{-i\w t} + \mathcal{O}\left(r^{-1} \right) ,   \hspace{4mm} \text{for}  \, l \geq 2.  \label{eqn:even_master_function_PW} 
\end{eqnarray}
where
\begin{equation} 
C_{l2}= \left(\frac{(2l+1)}{4\pi}\frac{(l-2)!}{(l+2)!}\right)^{1/2}.
\end{equation}
It follows from the reality condition on $h_{\mu \nu}$ that $\Phi_{l,-m,p}= \Phi^{*}_{lmp}$. Only the $m \pm 2$ and $l \ge 2$ modes are needed; all other modes ($l<2$ or $m\neq \pm2$) are zero for the plane wave (see Appendix \ref{App:plane_waves}).

\subsection{Scattering cross section\label{subsec:crosssection}}

Once the scattered radiation has been found via Eq.~(\ref{eq:asymp_matching_masters}), the associated energy flux at infinity can be calculated. The scattering cross section, $d \sigma/d \Omega$, is the energy flux per unit solid angle in the scattered radiation, divided by the energy flux per unit area in the incident plane wave. As shown in Appendix \ref{App:scat_flux} (see also Ref.~\cite{Futterman:1988ni}), the cross section can be written as the sum of the square modulus of a helicity-preserving scattering amplitude, $f(\theta)$, and a helicity-reversing amplitude, $g(\theta)$,
\begin{equation} \label{eq:total_cs}
\frac{d \sigma}{d \Omega}  =    |f(\theta)|^2 + |g(\theta)|^2,
\end{equation}
where 
\begin{eqnarray} \label{eq:sc_f}
f(\theta) &:=&  \frac{\pi}{\omega} \sum_{l,p} \left( \frac{(2l+1)}{4\pi} \right)^{1/2} \left( e^{2 i \delta_l^p} -1 \right) \,  _{-2}Y^{l2}(\theta), \\ \label{eq:sc_g}
g(\theta) &:=&  \frac{\pi}{\omega}  \sum_{l,p} p \left( \frac{(2l+1)}{4\pi} \right)^{1/2} \left( e^{2 i \delta_l^p} -1 \right) \, _{2}Y^{l2}(\theta).
\end{eqnarray}
The spherical harmonics with spin-weight $s$, $_s Y^{lm}(\theta,\phi)$, were introduced by Goldberg \cite{Goldberg:1966uu}, and we have suppressed the $\phi$ dependence. The phase shifts $ \delta_l^p$ are defined in terms of the mode coefficients $A_{lp}^{\text{in/out}}$ in Eq.~(\ref{eq:AinAout}) by
\begin{equation}
e^{2 i \delta_l^p} =  (-1)^{l+1} \frac{ A_{lp}^{\text{out}} }{ A_{lp}^{\text{in}} }.
\end{equation}
%
%

\subsection{Numerical method\label{subsec:nummethod}}

In Sec.~\ref{subsec:solution} we outlined how a regular solution satisfying physically-motivated boundary conditions could be constructed. Here, we give details of how we compute $A_{lp}^{\text{out}}$ and $A_{lp}^{\text{in}}$, and thus the scattering coefficients $e^{2 i \delta_l^p}$ in practice. 

The odd-parity master function for the interior perturbation has a regular Frobenius series solution at the origin 
\begin{equation}
\Qt \sim r^{l+1} \sum_{j=0}^{k} q_{2j} r^{2j}.
\end{equation}
The series coefficients $q_j$ can be found by inserting this series into Eq.~(\ref{eq:ODE_odd_interior}), and fixing the normalisation by choosing $q_0=1$. 

For the even parity system, the ODE system has regular singular points at $r=0$ and at $r=R$. We can find series solutions near the origin ($r_0 = 0$) and near the surface ($r_0 = R$) by expressing the equations in matrix form,
\begin{equation}\label{eqn:matx_polar}
\vec{Y'} = \frac{1}{z} \textbf{M} \cdot \vec{Y},
\end{equation}
where $\vec{Y} = [ \St,\Ft, z\St',z\Ft' ]^t$, 
$z = |r-r_0|$, $\vec{Y}'$ denotes $d\vec{Y}/dz$, and $\textbf{M}$ is a $4 \times 4$ matrix. One may then expand the matrix in a power series,
\begin{equation}\label{eqn:matx_expansion}
\textbf{M} =  \sum_{j=0}^{\infty} z^j \textbf{M}_j,
\end{equation}
where $\textbf{M}_j$ are constant matrices, and make the ansatz
\begin{equation}\label{eqn:vecsol_expansion}
\vec{Y} = z^{\sigma} \sum_{j=0}^{\infty} z^j \vec{Y}_j.
\end{equation}
Substituting Eqs.~(\ref{eqn:matx_expansion}) and (\ref{eqn:vecsol_expansion}) into Eq.~(\ref{eqn:matx_polar}), and equating coefficients of $z^{\sigma}$ gives 
\begin{subequations}
\begin{align}\label{eqn:eigenvalues_origin}
(\textbf{M}_0 - \sigma \textbf{I})\cdot  \vec{Y}_0 &= 0, \\
\label{eqn:higherorderseries_origin}
\left[ \textbf{M}_0 - \left( \sigma + k \right) \textbf{I} \right] \cdot \vec{Y}_k &= - \sum_{j=1}^{k} \textbf{M}_j \cdot \vec{Y}_{k-j}.
\end{align}
\end{subequations}
Equation (\ref{eqn:eigenvalues_origin}) determines the eigenvalues $\sigma$ and the corresponding eigenvectors $\vec{Y}_0$, and Eq.~(\ref{eqn:higherorderseries_origin}) generates the higher terms $\vec{Y}_j$ in the series solutions. At the origin, two of these series are regular, $\{ \vec{Y}^{(1)}, \vec{Y}^{(2)}\}$, with eigenvalues $\sigma=l+1$. Starting with initial conditions $\vec{Y}^{(1)}$, $\vec{Y}^{(2)}$ and $\tilde{Q}$ at $r=\epsilon$, where $\epsilon$ is some small value, we then numerically integrate the coupled ODEs (\ref{eq:ODE_odd_interior}), (\ref{eqn:interiorS}), and (\ref{eqn:interiorF}), to extend the solutions to $r=R$. Typically we use $\epsilon=10^{-6}R$ and we expand the series to order $k=15$. 

For the polytropes with $n\neq 0$, the speed of sound goes to zero at the stellar surface ($c_s \rightarrow 0$ as $r \rightarrow R^-$), and Eq.~(\ref{eqn:interiorF}) cannot be used reliably near $r=R$. Keeping only the terms of order $1/c_s^2$ in Eq.~(\ref{eqn:interiorF}) allows us to solve for $\vec{Y}_j$ between $r=R-\epsilon$ and $r=R$. Imposing the boundary condition and utilising the junction conditions gives the odd and even parity master functions, $\tilde{\Phi}_{lm}^{\pm}(R,\omega)$ , at the (outer) surface (see Section \ref{sec:perts_surface}). We integrate these out to some large value of $r=r_{max}$, typically choosing $r_{max} \approx 100 R$. 

In the far field we compute generalised series solutions for $\tilde{\Phi}_{lm}^{\pm}$ about $r=\infty$ of the form
\begin{equation}
\tilde{\Phi}_{lmp}^{out}(r,\omega) \sim  e^{-i\omega r_*} \sum_{j=0}^{N} b_j r^{-j}, \hspace{5mm} \tilde{\Phi}_{lmp}^{in}(r,\omega) = \tilde{\Phi}_{lmp}^{out*}(r,\omega).
\end{equation}
%
We choose $N=15$ to achieve accurate results. We then match this to the numerical solution to obtain the mode coefficients by solving 
\vspace{.2cm}
\begin{equation}
\begin{pmatrix}
\tilde{\Phi}_{lmp}^{out} & \tilde{\Phi}_{lmp}^{in} \\
\p_r \tilde{\Phi}_{lmp}^{out} & \p_r \tilde{\Phi}_{lmp}^{in}
\end{pmatrix} \begin{pmatrix}
A_{lp}^{out} \\ A_{lp}^{in} 
\end{pmatrix} = \begin{pmatrix}
\tilde{\Phi}_{lmp} \\ \p_r \tilde{\Phi}_{lmp}
\end{pmatrix} \Bigg|_{r=r_{max}}.
\end{equation}

\section{Results\label{sec:results}}

The scattering process is encapsulated by the specific stellar model or compact object used, and the dimensionless parameters 
$
M \omega = \frac{\pi R_S}{\lambda}$ and $\frac{R}{M} = \frac{2 R}{R_S},
$
where $R_S =2 M$ is the Schwarzschild radius and $\lambda$ is the wavelength of the incident wave. We consider a range of couplings $M \omega \sim 0.1-10$, and the tenuity $R/M = 6$, which is comparable to that of a neutron star. After fixing the tenuity, the polytropic index $n$ determines the stellar structure in our model. 

\begin{figure}
\centering
\includegraphics[width=.8\textwidth]{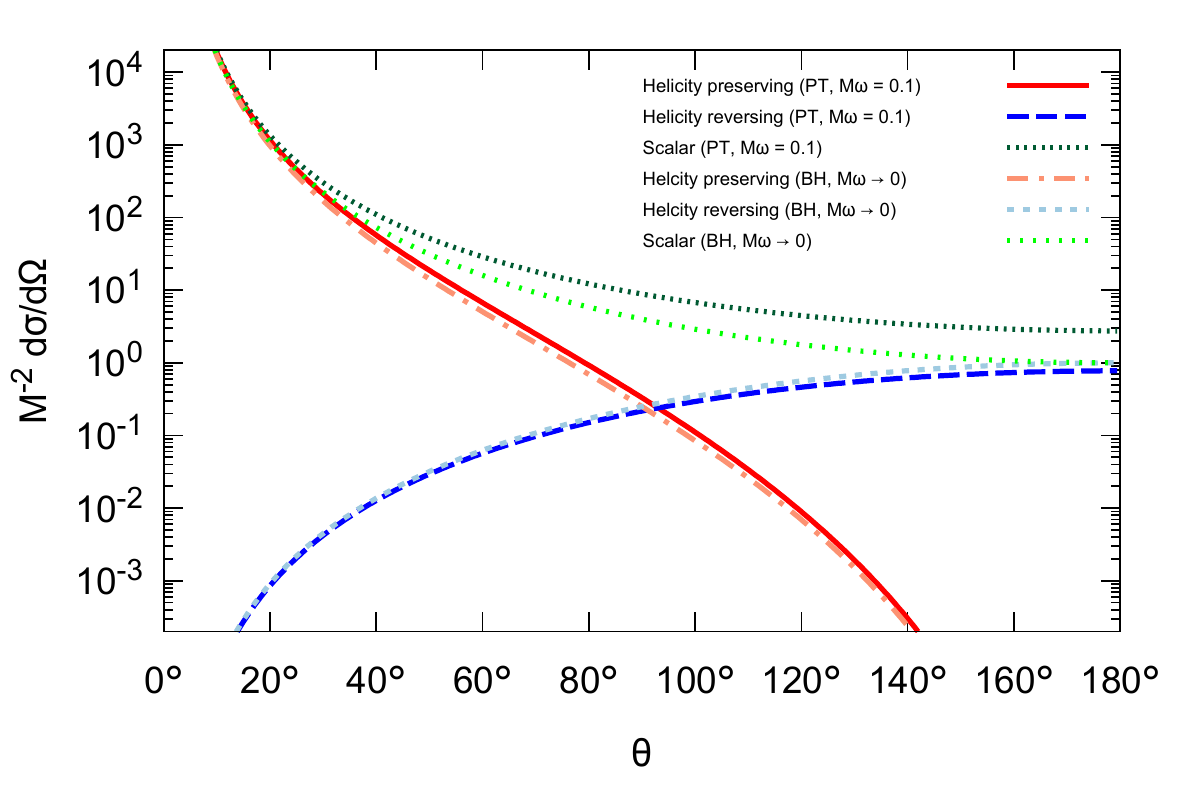}
\caption{Helicity preserving (red) and helicity reversing (blue) gravitational scattering cross sections for a polytropic (PT) star with tenuity $R/M=6$, polytropic index $n=1$, and coupling $M\omega=0.1$. The cross section for a scalar wave incident on the star is also shown (green). The low-frequency approximations for black hole (BH) scattering cross sections, given in Eq.~(\ref{eq:lowfreq_grav_scat}), are shown as dotted lines.  }
\label{fig:lowfreqappx}
\end{figure}

Low-frequency ($M \omega \ll 1$) analytic approximations for scattering by a black hole are summarised in Eq.~(\ref{eq:lowfreq_grav_scat}). In Fig.~\ref{fig:lowfreqappx} we compare the approximations with numerically-determined scattering cross sections for an $n=1$, $R/M=6$ polytrope with $M\omega=0.1$. We find that the cross sections are very similar, and the polytrope cross sections appear to approach the low frequency black hole approximations as $M\omega \rightarrow 0$. This is consistent with the interpretation that long wavelength waves do not `see' the strong-field structure of the scatterer if $\lambda \gg R$, and thus the cross section is insensitive to the nature of the central body. 

\begin{figure}
\centering
\includegraphics[width=.8\textwidth]{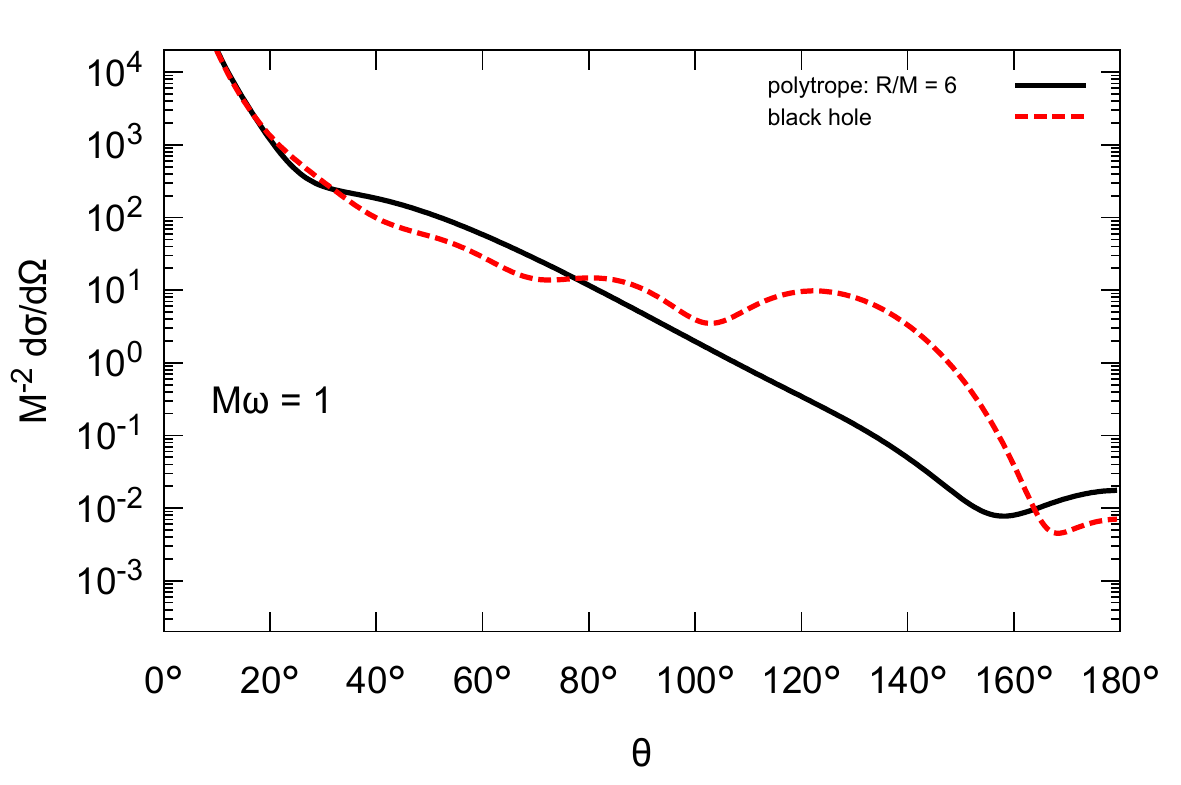}
\caption{Gravitational wave scattering cross sections for a polytropic star with $R/M=6$, polytropic index $n=1$ (black, solid) and a Schwarzschild black hole (red, dashed). The coupling is $M \omega =1$. }
\label{fig:PT_BH_Mw1}
\end{figure}

\begin{figure}
\centering
\subfloat[]{\label{subfig:csec:wpt1}%
  \includegraphics[width=0.5\textwidth]{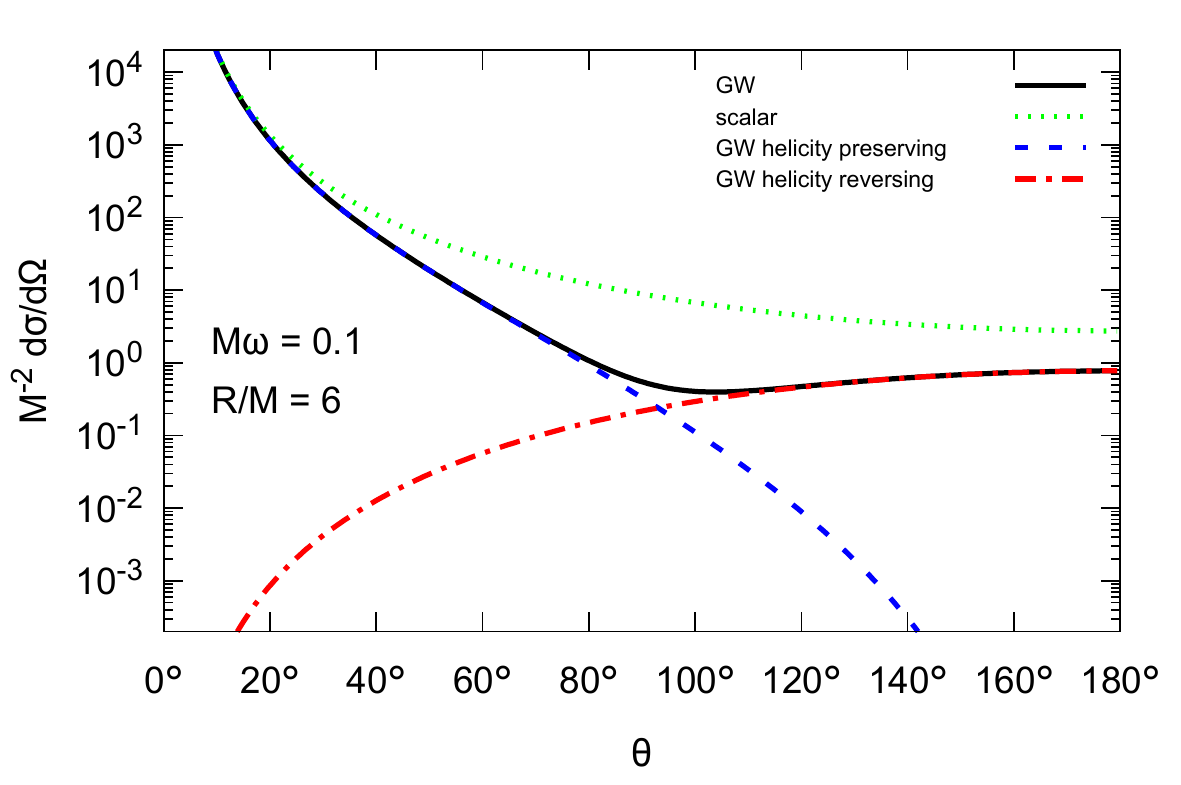}
} 
\subfloat[]{\label{subfig:csec:w1}%
  \includegraphics[width=0.5\textwidth]{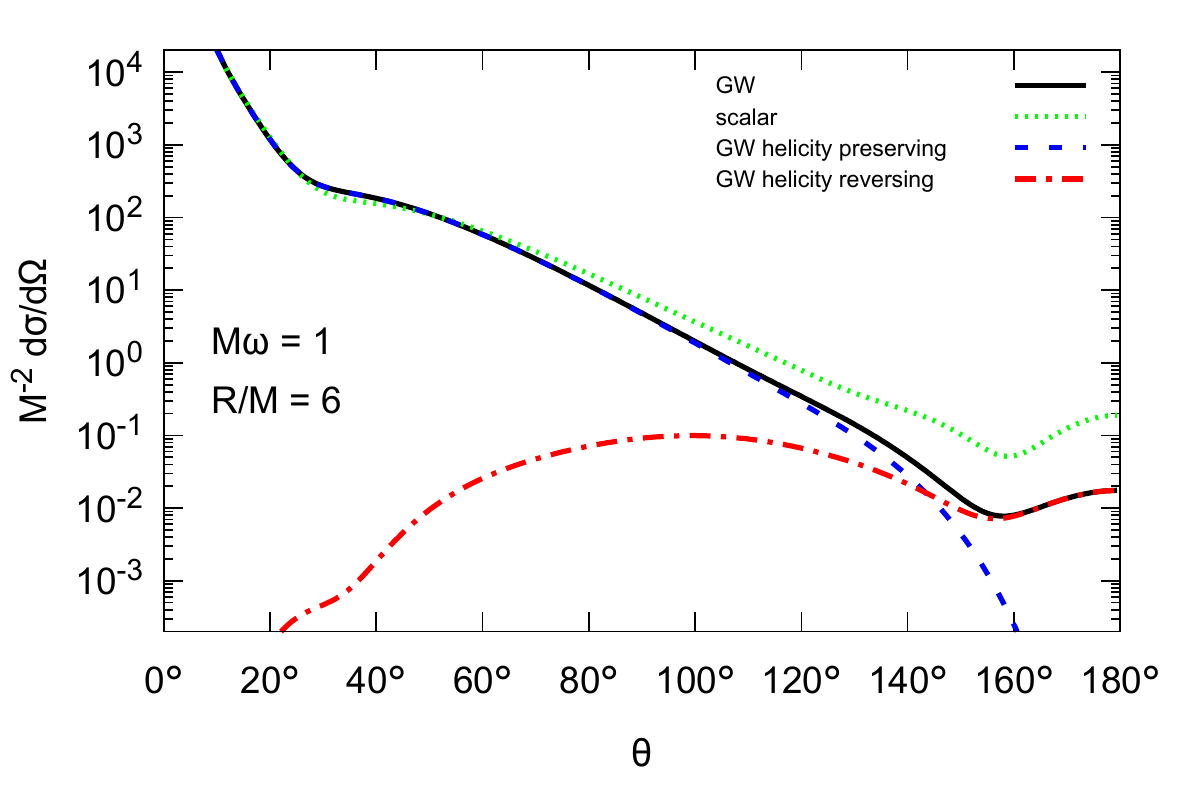}
} \\
\subfloat[]{\label{subfig:csec:w2}%
  \includegraphics[width=0.5\textwidth]{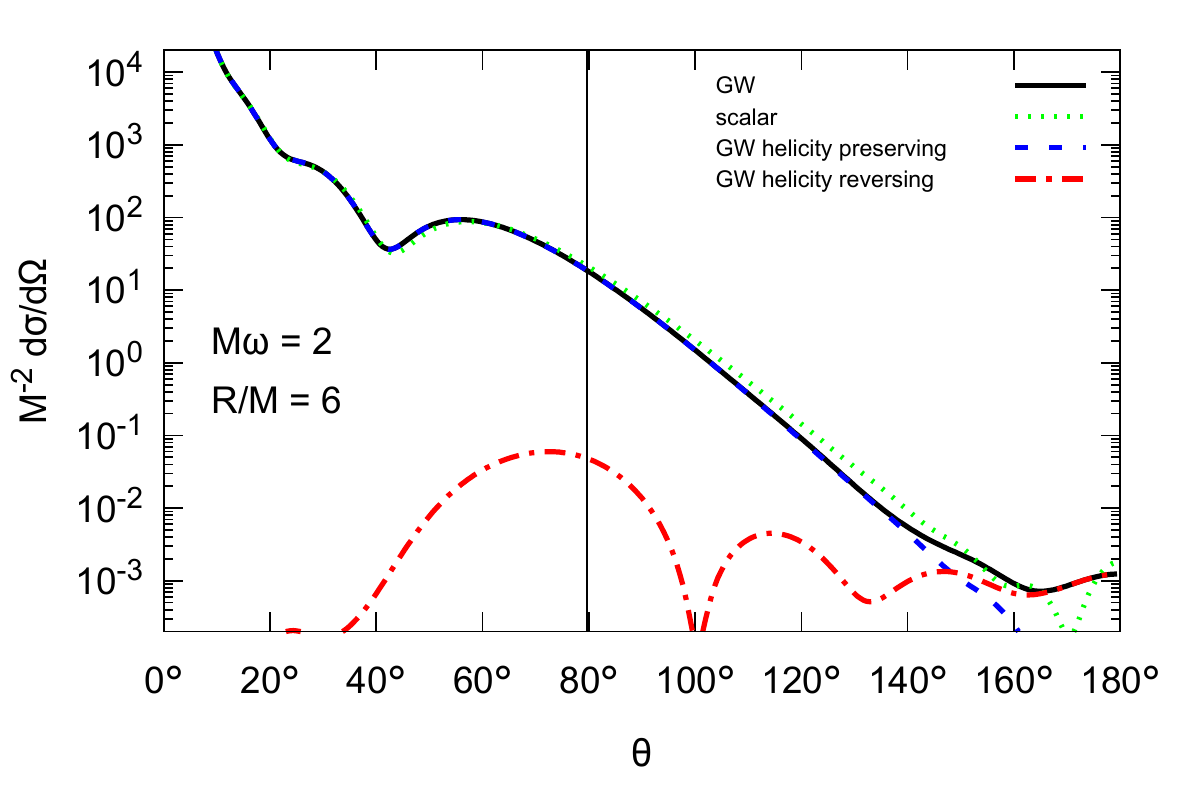}
}
\subfloat[]{\label{subfig:csec:w4}%
  \includegraphics[width=0.5\textwidth]{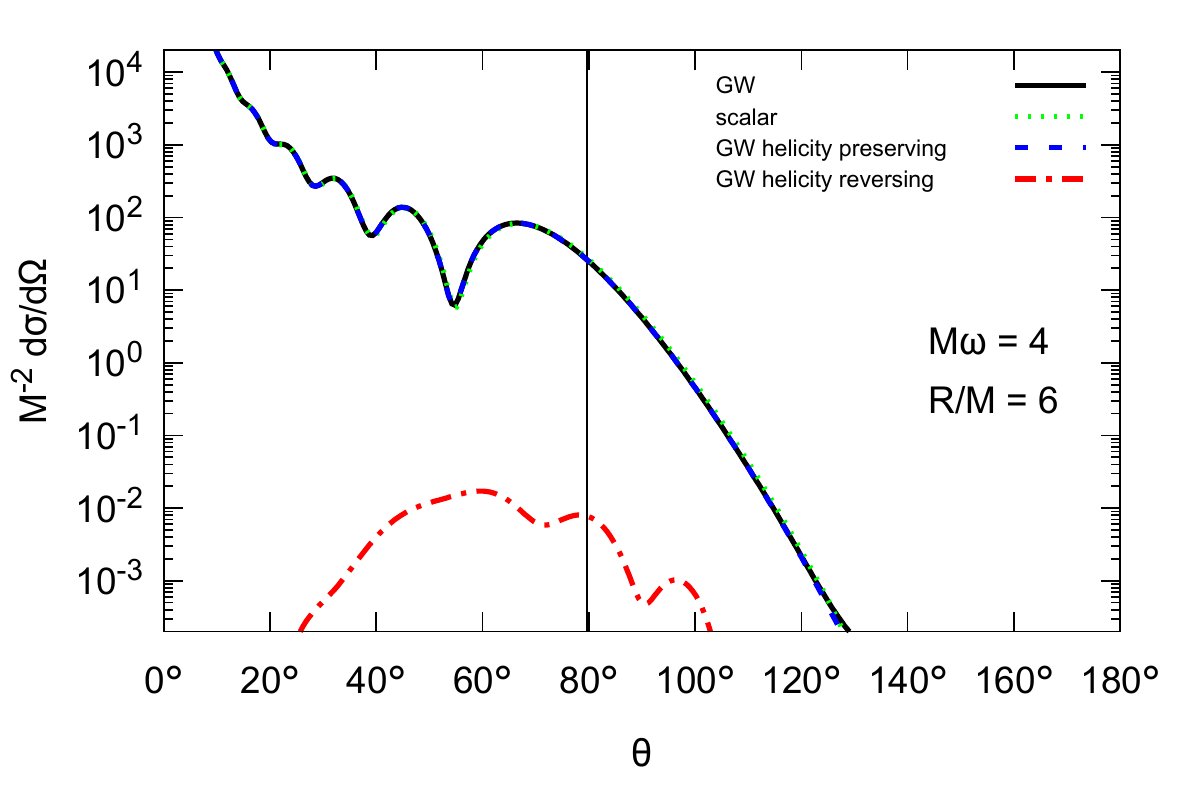}
}
\caption{
Scattering cross sections for a polytropic star with $R/M=6$, polytropic index $n=1$ and coupling (a) $M \omega =0.1$, (b) $M \omega =1$, (c) $M \omega =2$ and (d) $M \omega =4$. The helicity preserving (reversing) part of the GW cross section is shown in dashed blue (dot-dashed red), and the cross section for a scalar wave is shown in dotted green. The rainbow angle, $\theta_{r} \approx 79.7 ^{\circ}$, is shown as a solid vertical line for the two higher frequency cases. 
}
\label{fig:n1_panel}
\end{figure}

The universality of the cross section at low $M \omega$, seen in Fig.~\ref{fig:n1_panel}, does not persist at higher frequencies. Figure \ref{fig:PT_BH_Mw1} shows the case $M \omega = 1$, where the cross sections for GWs scattering from a compact body ($R/M = 6$) are clearly different from those for scattering from a black hole, with visible differences occurring at large angles ($\theta \gtrsim 20^\circ$). The differences become more marked at shorter wavelengths (higher frequencies), as the wave can resolve and probe the details of the internal structure of the body. 

Figure \ref{fig:n1_panel} shows the cross sections for a polytropic star with $n=1$, $R/M=6$, and a range of couplings $M\omega$. 
At low $M \omega$, the helicity-reversing cross section $|g|^2$ plays a role in large-angle scattering. As shown in plots (b), (c) and (d), the contributions from $|g|^2$ diminishes at higher $M\omega$, and it is negligible for $M\omega = 4$.  
Once the coupling is sufficiently large, $M \omega \gtrsim 2$, a rainbow pattern appears in the cross section. That is, a primary peak at some $\theta_p $, which may be followed by supernumerary troughs and peaks at $\theta < \theta_p $, and a falling off of the cross section into the `shadow zone', $\theta > \theta_p $. The primary peak is close to the rainbow angle, $\theta_r$, which is the maximum deflection angle for a null geodesic incident on the compact body (for details of how $\theta_r $ is calculated see \cite{Dolan:2017rtj}). The rainbow feature is superposed on a Rutherford-type scattering cross section, with a forward divergence, which essentially arises because we are modelling a plane wave of infinite extent in a long-range field.  In the semiclassical regime, $M\omega \gg 1$, $\theta_p $ approaches $\theta_r$ from below, and the width of the oscillations in the cross section decreases, as may be anticipated from Airy's formula, Eq.~(\ref{eq:airy}). 

\begin{figure}
\centering
\includegraphics[width=.8\textwidth]{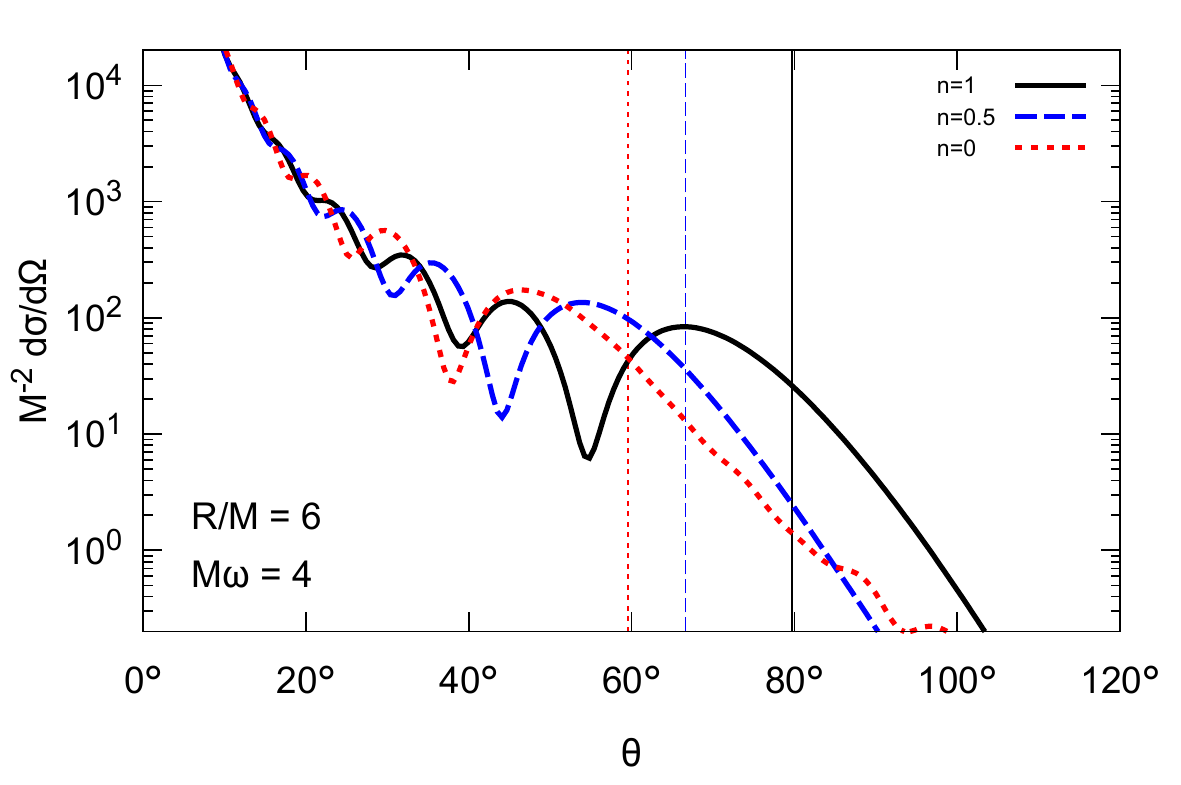}
\caption{Rainbow scattering for three polytropes with $n=1$ (black, solid), $n=0.5$ (blue, long-dashed), and $n=0$ (red, short-dashed). For smaller $n$ the rainbow angle, and thus the positions of the primary and supernumerary troughs and peaks, are shifted to smaller angles. The rainbow angles are indicated with vertical lines with the same style as the corresponding cross section. The rainbow angle is $\theta_r \approx 59.6^{\circ}, 66.6^{\circ},$ and $79.7^{\circ}$ for $n=0, 1,$ and $2$ respectively. The generic rainbow scattering pattern, superimposed on a divergence at $\theta=0$, remains for all cases.}
\label{fig:rainbow_dn}
\end{figure}

We find that rainbow scattering is seen for a range of stellar models at higher frequencies. Figure \ref{fig:rainbow_dn} shows the scattering cross sections for polytropes with polytropic index $n=0,1,2$, with $n=0$ corresponding to a star of constant density. Generally, as $n$ increases, the bodies mass becomes more concentrated in the centre, and the star becomes `less stiff', with a slower internal speed of sound. As a consequence, the maximally-deflected geodesic, which passes through the body, scatters through a greater angle. This is confirmed for the small sample of polytropes we consider in Table \ref{Tab:bow_angles}. Consequently, the rainbow pattern of supernumerary peaks and troughs is shifted to higher angles for larger $n$, as shown in Figure \ref{fig:rainbow_dn}. The stellar structure is clearly affecting the scattering cross section. The inverse problem of determining stellar structure from a scattering cross section would be worth addressing, in principle. 

\begin{table}[]
\centering
\begin{tabular}{ | c | c | c | c |}
 \hline
 \diagbox
 {$R/M$}{$n$} & $0$ & $0.5$ & $1$   \\ 
 \hline
 $5$ & $81.1^{\circ}$ & $92.1^{\circ}$ & $115^{\circ}$ \\
 \hline
 $6$ & $59.6^{\circ}$ & $66.6^{\circ}$ & $79.7^{\circ}$ \\
 \hline
 $7$ & $47.3^{\circ}$ & $52.4^{\circ}$ & $61.5^{\circ}$ \\
 \hline
\end{tabular}
\caption{Rainbow angles, $\theta_r$, for polytropes with tenuity $R/M$ and polytropic index $n$.}
\label{Tab:bow_angles}
\end{table}

\section{Discussion and conclusions\label{sec:conclusions}}
In the preceding sections, we have presented numerical results for GW scattering cross sections for compact bodies of tenuity $R/M = 6$, modelled as spherically-symmetric polytropes. We may now draw several conclusions. 

\begin{enumerate}
 \item The cross section for the gravitational wave is qualitatively similar to that for a scalar field $\Phi$ which is not directly coupled to the matter sector (studied in Paper I \cite{Dolan:2017rtj}), with some minor differences at lower frequencies.
 \item At low frequencies (long wavelengths) $M \omega \ll 1$, the scattering cross section is insensitive to the internal structure of the compact body, and the cross section for a compact body reduces to that for a Schwarzschild black hole of the same mass, given by Eq.~(\ref{eq:lowfreq_grav_scat}). At low frequencies and large angles $\theta \gtrsim 90^\circ$ there is a significant contribution from the helicity-reversing amplitude $|g|^2$, due to the fact that the phase shift for odd and even-parity perturbations differs (see Fig.~\ref{fig:lowfreqappx}). 
 \item The contribution from the helicity-reversing amplitude diminishes as the frequency increases, becoming negligible in practice for $M\omega \gtrsim 1$ (see Fig.~\ref{fig:n1_panel}). 
 \item The even-parity perturbation in the gravitational wave couples directly to the fluid degrees of freedom of the star, whereas the odd-parity part does not. However, the lack of a significant helicity-reversing amplitude, implying that the odd and even-parity phase shifts are approximately equal, would appear to imply that this coupling plays no major role in time-independent scattering. (Fluid motions could be important in the time-dependent context, as suggested by Allen \emph{et al.}~\cite{Allen:1997xj}.) 
 \item At high frequencies, the gravitational wave exhibits the expected features of rainbow scattering (see Figs.~\ref{fig:n1_panel} and \ref{fig:rainbow_dn}), as anticipated from the ray analysis in Fig.~\ref{fig:cuspcaustic}--\ref{fig:NF_scalarMw8}, Eq.~(\ref{eq:airy}) and Paper I.
 \item A wavefront passing through a neutron star will be focussed at a cusp caustic (see Figs.~\ref{fig:cuspcaustic} and \ref{fig:NF_scalarMw8}). For tenuity $R/M  = 6$ the cusp caustic may form inside the star, or just outside, depending on the polytropic index $n$. 
 \item The rainbow angle $\theta_r$ is sensitive to both the polytropic index $n$ and tenuity $R/M$ (see Fig.~\ref{fig:rainbow_dn} and Table \ref{Tab:bow_angles}).
\end{enumerate}
The direct coupling between the gravitational wave and fluid motions appears to be inconsequential in time-independent scattering. On the other hand, the internal structure of the fluid star, determined by its EoS, is certainly not. Constraining the EoS for a neutron star is an important goal where multiple disciplines overlap including nuclear, particle and gravitational physics. The EoS determines the body's density and pressure profiles, which consequently alter the space-time curvature, the effective potential and thus the scattering of the incident wave. We have shown here that the rainbow angle, and thus the position of the primary rainbow maximum for moderate-to-high frequencies, is sensitive to the EoS.

Rainbow scattering is a wave phenomenon that arises on disparate scales in physics. In the context of ion-scattering experiments \cite{Goldberg:1972zzb, Goldberg:1974zza, Khoa:2006id}, rainbow scattering was used to discriminate between competing models of the nuclear potential (see e.g.~Fig.~11 in \cite{Khoa:2006id}).  In fact, in the nuclear case the quantum-mechanical deflection function possesses \emph{two} stationary points, linked to the (repulsive) long-ranged Coulomb interaction and the (attractive) short-range nuclear interaction, respectively: see Fig.~9 in \cite{Khoa:2006id}. The former leads to small-angle rainbow scattering and the latter to the wide-angle rainbow features from which one may infer basic properties of the nuclear potential.

Let us now consider gravitational wave scattering by neutron stars in an astrophysical context. 
A typical neutron star has a mass of $M \approx 1.5 M_{\odot}$, corresponding to a gravitational radius of $r_g = GM/c^2 \approx 2.2$\,km. For this mass, the range of frequencies considered here, $M \omega \sim 0.1-10$ in geometric units, corresponds to wave frequencies in the range $\omega \sim 10^4$--$10^6$ Hz. Known astrophysical sources emit GWs with frequencies $\omega \lesssim 10^3$ Hz \cite{Moore:2014lga}, with the highest frequency source being millisecond pulsars. (The fastest-spinning pulsar known, PSR J1748-2446ad, has a frequency of $716$\, \text{Hz}. If it is not axisymmetric about its rotation axis, it will emit gravitational waves \cite{Authors:2019ztc,Abbott:2019bed} with $\omega = 9 \times 10^3$\, Hz). Thus the low-frequency approximation, Eq.~(\ref{eq:lowfreq_grav_scat}), is likely to be approximately valid for generic time-independent scattering scenarios involving gravitational waves and neutron stars. At low frequencies, the cross section is insensitive to the internal structure of the star, and so rainbow scattering is not manifest. 

The conditions for rainbow scattering will arise naturally for gravitational waves impinging upon larger, less compact bodies such as white dwarfs ($R/M \sim 1400$); 
or upon intermediate-mass black holes ($10^2$--$10^5 M_\odot$) and supermassive black holes $M \gtrsim 10^5 M_\odot$ surrounded by a shell of matter \cite{Leite:2019uql}. 



The rainbow interference effect can arise naturally, too, for other weakly-interacting fields. Motivated by neutrino oscillations, Alexandre and Clough recently investigated the plane wave scattering of coupled and flavoured massive scalars on a Schwarzschild black hole background \cite{Alexandre:2018crg}. They showed that a long-range interference pattern will form, altering the flavour oscillation probability. They postulate that this effect may also be seen for neutrinos, and that unexpected neutrino detection patterns could be observed when a black hole is situated between a terrestrial detector and neutrino source. The consequences of replacing the black hole with a dense compact body are yet to be explored fully. For example, it is not known if the rainbow scattering cross sections for neutron stars can be accurately calculated via a sum of Regge poles (see Ref.~\cite{Folacci:2019cmc} for recent progress in the black hole context). 

An open question is whether a rainbow from (quasi-)time independent scattering could ever be detected in practice, in the gravitational context. To verify the diffraction pattern, a detector would need to sample at least one peak and trough of the interference pattern. This would certainly require a fortuitous alignment of scatterer and source, such that a detector lay just inside the rainbow angle $\theta_r$. As the angle of observation $\theta_{\text{obs}}$ would be essentially fixed, the detector would need to sample across a range of frequencies $\Delta \omega$. This would necessitate a gravitational wave source that is either multiband, like an eccentric binary, or which sweeps across a range of frequencies over time, like a binary inspiral. The Airy formula (\ref{eq:airy}) yields an estimate $\frac{\Delta \omega}{\omega} \sim \frac{2 q}{\theta_r - \theta_{\text{obs}}}$ for the range of frequencies that would be needed to sample the rainbow.

One interesting avenue for further work is to consider the possible physical consequences of the cusp caustic that forms in the gravitationally-scattered wavefront (Fig.~\ref{fig:cuspcaustic}), due to the focussing effect of gravity. The position of the cusp caustic is sensitive to the density and compactness of the compact body. Whereas for a neutron star it can form inside the star, or close to its crust, for a dilute body like the Sun it will form at a distance of approximately 550 astronomical units \cite{Bontz:1981}. Cusp caustics also arise naturally in other wave-propagation contexts. For example, in shallow water, a wave propagates more slowly wherever it passes over a submerged island. Berry  \cite{Berry:2005:tsunami,Berry:2007:focused} has described how cusp caustics may arise in tsunamis due to seabed topography, leading to the focussing of energy, with potentially devastating consequences. There are close parallels between shallow water wave propagation and gravitational waves on a curved spacetime which we may explore in future work.

\acknowledgments
T.S.~acknowledges financial support from EPSRC. S.R.D.~acknowledges financial support from the European Union's Horizon 2020 research and innovation programme under the H2020-MSCA-RISE-2017 Grant No.~FunFiCO-777740, and from the Science and Technology Facilities Council (STFC) under Grant No.~ST/P000800/1.

\appendix

\section{Plane wave matching}\label{App:plane_waves}

The scattered part of the metric perturbation is found by matching to the total metric perturbation and the plane wave via
\begin{equation} \label{eq:asymp_matching}
h_{AB}^{scat} = h_{AB}^{total}  - h_{AB}^{plane}.
\end{equation}
In order to carry this subtraction out in a gauge invariant way we instead use the master functions, as in Eq.~(\ref{eq:asymp_matching_masters}). We assume that the scattered part is all outgoing in the far field
\begin{equation}
\tilde{\Phi}_{lmp}^{scat}(r,\omega) \sim \alpha_{lmp}(\omega)  e^{-i \omega r_*}.
\end{equation}
The leading order behaviour of the scalar master functions for a left circularly polarized distorted plane wave in the far field is given by Eqs.~(\ref{eqn:odd_master_function_PW}) and (\ref{eqn:even_master_function_PW}). The method for finding the master functions of a given perturbation is described in Martel and Poisson (MPo) \cite{Martel:2005ir}, and Barack and Lousto (BL) \cite{Barack:2005nr}. Table III of BL provides the necessary differential operations to apply to $h_{\mu\nu}$ to find the coefficients of their particular spherical harmonic decomposition. The decomposition of MPo is similar to that of BL, so it is simple to then deduce the decomposition coefficients of MPo that define the master functions in their Eqs.~(4.23) and (5.13). Substituting Eqs.~(\ref{eq:bc_farfield}), (\ref{eqn:odd_master_function_PW}) and (\ref{eqn:even_master_function_PW}) into Eq.~(\ref{eq:asymp_matching_masters}), gives  
\begin{eqnarray}
\alpha_{lm-}(\omega) &=& \phantom{-i} \frac{(2 \pi)^2 H C_{l2}}{\omega} \left( e^{2 i \delta_l^-(\omega)} -1 \right) \left[ \delta_{m2} \delta(\w -\omega) - \delta_{m,-2} \delta(\w +\omega) \right], \\
\alpha_{lm+}(\omega) &=& -  \frac{(2 \pi)^2  H C_{l2} } {\omega} i \left( e^{2 i \delta_l^-(\omega)} -1 \right) \left[ \delta_{m2} \delta(\w -\omega) + \delta_{m,-2} \delta(\w +\omega) \right].
\end{eqnarray}
The metric perturbation corresponding to the scattered wave in the far field can then be reconstructed with Eq.~(\ref{eqn:total_rad_pert}),
\begin{align} \label{eqn:total_scattered_pert}
h_{AB}^{scat}  \sim  \frac{r}{2\pi} \int_{-\infty}^{\infty} d \omega \left[ \sum_{l,m,p}    \alpha_{lmp}(\omega) e^{i \omega (t- r_*)} X_{AB}^{lmp} \right] .
\end{align}
It is convenient to be able to switch to using spin-weighted spherical harmonics, $_s Y^{lm}$, defined in \cite{Goldberg:1966uu}, where $s$ is the spin weight. They satisfy $_s \bar{Y}^{lm} = (-1)^{m+s} \, _{(-s)}Y^{l(-m)}$, where an over-bar denotes the complex conjugate. The MPo harmonics can be written in terms of spin-weight $\pm2$ spherical harmonics,
\begin{eqnarray} \label{eqn:MP_spinweighted}
X_{AB}^{lm-} &=& \frac{i}{2} \sqrt{(l-1)l(l+1)(l+2)} \left( \, _{2}Y^{lm} \bar{m}_A \bar{m}_B - \, _{-2}Y^{lm} m_A m_B   \right), \\
X_{AB}^{lm+} &=& \frac{1}{2} \sqrt{(l-1)l(l+1)(l+2)} \left( \, _{2}Y^{lm} \bar{m}_A \bar{m}_B + \, _{-2}Y^{lm} m_A m_B   \right),
\end{eqnarray}
where $m_A=2^{-1/2}(1, i \sin \theta)$. Note $X_{AB}^{l(-m)p}= (-1)^m \bar{X}_{AB}^{lmp}$. 
%

\section{Scattered flux}\label{App:scat_flux}
 
The energy flux per unit solid angle in the scattered radiation is given by (pg 72 \cite{Futterman:1988ni}),
\begin{equation}
\frac{dE}{d t \, d \Omega} = \lim_{r \rightarrow \infty} r^2 \tensor{T}{^{r}_{t}}.
\end{equation}
A gauge invariant way to calculate the stress energy of a metric perturbation is to use a space-time averaging process \cite{Isaacson:1968zza,Brill:1964zz}, denoted by placing angular brackets around the quantity to be averaged, $\langle \cdot \cdot \cdot \rangle$. The Brill-Hartle (BH) averaged stress energy tensor of a metric perturbation is
\begin{equation}
\left\langle T_{\mu\nu} \right\rangle = \frac{1}{32 \pi} \left\langle \tensor{h}{^{\rho\tau}_{;\mu}} \tensor{h}{_{\rho\tau}_{;\nu}}  \right\rangle.
\end{equation}
The radiative part of the relevant component of the stress energy tensor is
\begin{equation}\label{eqn:BH_Ttensor}
\left\langle \tensor{T}{^{r}_{t}} \right \rangle \sim \frac{1}{r^4} \frac{1}{32\pi} \left\langle \p_r h^{AB} \p_t h_{AB} \right\rangle \hspace{5mm} \text{as	} r \rightarrow \infty.
\end{equation}
Substituting Eq.~(\ref{eqn:total_scattered_pert}) into Eq.~(\ref{eqn:BH_Ttensor}), expressing the MPo spherical harmonics in terms of spin weighted spherical harmonics (Eq.~(\ref{eqn:MP_spinweighted})), and performing the BH averaging over a region much larger than the wavelength, results in 
\begin{eqnarray} \label{eq:flux}
\left< \frac{dE}{d t \, d \Omega} \right> = \frac{H^2}{16 \pi } \bigg( \Big| &\pi & \sum_{l,p} \left( \frac{(2l+1)}{4\pi} \right)^{1/2} \left( e^{2 i \delta_l^p( \omega_0)} -1 \right) \,  _{-2}Y^{l2}(\theta) \Big|^2  \nonumber \\ 
  &+& \Big| \pi \sum_{l,p} p \left( \frac{(2l+1)}{4\pi} \right)^{1/2}  \left( e^{2 i \delta_l^p( \omega_0)} -1 \right) \,  _{2}Y^{l2}(\theta) \Big|^2 \bigg).
\end{eqnarray}
The flux per unit area in the incident plane wave is 
\begin{equation} \label{eq:plane_wave_flux}
\frac{dE}{dt \, dA}\bigg|_{Plane} = \frac{H^2 \omega^2}{16 \pi}. 
\end{equation}
The scattering cross section is defined as 
\begin{equation} \label{eq:scat_cs_def}
\frac{d \sigma }{d \Omega} := \frac{dE}{d t \, d \Omega} \bigg/ \frac{dE}{dt \, dA}\bigg|_{Plane}.
\end{equation}
Substituting Eq.~(\ref{eq:flux}) and (\ref{eq:plane_wave_flux}) into Eqs.~(\ref{eq:scat_cs_def}) gives Eq.~(\ref{eq:total_cs}), the scattering cross section for gravitational plane wave scattering.

%
%
%
%
%
%
%

\bibliography{scatteringCOGWbib}

\begin{thebibliography}{71}%
\makeatletter
\providecommand \@ifxundefined [1]{%
 \@ifx{#1\undefined}
}%
\providecommand \@ifnum [1]{%
 \ifnum #1\expandafter \@firstoftwo
 \else \expandafter \@secondoftwo
 \fi
}%
\providecommand \@ifx [1]{%
 \ifx #1\expandafter \@firstoftwo
 \else \expandafter \@secondoftwo
 \fi
}%
\providecommand \natexlab [1]{#1}%
\providecommand \enquote  [1]{``#1''}%
\providecommand \bibnamefont  [1]{#1}%
\providecommand \bibfnamefont [1]{#1}%
\providecommand \citenamefont [1]{#1}%
\providecommand \href@noop [0]{\@secondoftwo}%
\providecommand \href [0]{\begingroup \@sanitize@url \@href}%
\providecommand \@href[1]{\@@startlink{#1}\@@href}%
\providecommand \@@href[1]{\endgroup#1\@@endlink}%
\providecommand \@sanitize@url [0]{\catcode `\\12\catcode `\$12\catcode
  `\&12\catcode `\#12\catcode `\^12\catcode `\_12\catcode `\%12\relax}%
\providecommand \@@startlink[1]{}%
\providecommand \@@endlink[0]{}%
\providecommand \url  [0]{\begingroup\@sanitize@url \@url }%
\providecommand \@url [1]{\endgroup\@href {#1}{\urlprefix }}%
\providecommand \urlprefix  [0]{URL }%
\providecommand \Eprint [0]{\href }%
\providecommand \doibase [0]{http://dx.doi.org/}%
\providecommand \selectlanguage [0]{\@gobble}%
\providecommand \bibinfo  [0]{\@secondoftwo}%
\providecommand \bibfield  [0]{\@secondoftwo}%
\providecommand \translation [1]{[#1]}%
\providecommand \BibitemOpen [0]{}%
\providecommand \bibitemStop [0]{}%
\providecommand \bibitemNoStop [0]{.\EOS\space}%
\providecommand \EOS [0]{\spacefactor3000\relax}%
\providecommand \BibitemShut  [1]{\csname bibitem#1\endcsname}%
\let\auto@bib@innerbib\@empty
\bibitem [{\citenamefont {Abbott}\ \emph
  {et~al.}(2016{\natexlab{a}})\citenamefont {Abbott} \emph
  {et~al.}}]{Abbott:2016blz}%
  \BibitemOpen
  \bibfield  {author} {\bibinfo {author} {\bibfnamefont {B.~P.}\ \bibnamefont
  {Abbott}} \emph {et~al.} (\bibinfo {collaboration} {Virgo, LIGO
  Scientific}),\ }\href {\doibase 10.1103/PhysRevLett.116.061102} {\bibfield
  {journal} {\bibinfo  {journal} {Phys. Rev. Lett.}\ }\textbf {\bibinfo
  {volume} {116}},\ \bibinfo {pages} {061102} (\bibinfo {year}
  {2016}{\natexlab{a}})},\ \Eprint {http://arxiv.org/abs/1602.03837}
  {arXiv:1602.03837 [gr-qc]} \BibitemShut {NoStop}%
\bibitem [{\citenamefont {Abbott}\ \emph {et~al.}(2018)\citenamefont {Abbott}
  \emph {et~al.}}]{LIGOScientific:2018mvr}%
  \BibitemOpen
  \bibfield  {author} {\bibinfo {author} {\bibfnamefont {B.~P.}\ \bibnamefont
  {Abbott}} \emph {et~al.} (\bibinfo {collaboration} {LIGO Scientific,
  Virgo}),\ }\href@noop {} {\  (\bibinfo {year} {2018})},\ \Eprint
  {http://arxiv.org/abs/1811.12907} {arXiv:1811.12907 [astro-ph.HE]}
  \BibitemShut {NoStop}%
\bibitem [{\citenamefont {Abbott}\ \emph
  {et~al.}(2016{\natexlab{b}})\citenamefont {Abbott} \emph
  {et~al.}}]{Abbott:2016nmj}%
  \BibitemOpen
  \bibfield  {author} {\bibinfo {author} {\bibfnamefont {B.~P.}\ \bibnamefont
  {Abbott}} \emph {et~al.} (\bibinfo {collaboration} {Virgo, LIGO
  Scientific}),\ }\href {\doibase 10.1103/PhysRevLett.116.241103} {\bibfield
  {journal} {\bibinfo  {journal} {Phys. Rev. Lett.}\ }\textbf {\bibinfo
  {volume} {116}},\ \bibinfo {pages} {241103} (\bibinfo {year}
  {2016}{\natexlab{b}})},\ \Eprint {http://arxiv.org/abs/1606.04855}
  {arXiv:1606.04855 [gr-qc]} \BibitemShut {NoStop}%
\bibitem [{\citenamefont {Abbott}\ \emph
  {et~al.}(2017{\natexlab{a}})\citenamefont {Abbott} \emph
  {et~al.}}]{Abbott:2017vtc}%
  \BibitemOpen
  \bibfield  {author} {\bibinfo {author} {\bibfnamefont {B.~P.}\ \bibnamefont
  {Abbott}} \emph {et~al.} (\bibinfo {collaboration} {VIRGO, LIGO
  Scientific}),\ }\href {\doibase 10.1103/PhysRevLett.118.221101,
  10.1103/PhysRevLett.121.129901} {\bibfield  {journal} {\bibinfo  {journal}
  {Phys. Rev. Lett.}\ }\textbf {\bibinfo {volume} {118}},\ \bibinfo {pages}
  {221101} (\bibinfo {year} {2017}{\natexlab{a}})},\ \bibinfo {note} {[Erratum:
  Phys. Rev. Lett.121,no.12,129901(2018)]},\ \Eprint
  {http://arxiv.org/abs/1706.01812} {arXiv:1706.01812 [gr-qc]} \BibitemShut
  {NoStop}%
\bibitem [{\citenamefont {Abbott}\ \emph
  {et~al.}(2017{\natexlab{b}})\citenamefont {Abbott} \emph
  {et~al.}}]{Abbott:2017gyy}%
  \BibitemOpen
  \bibfield  {author} {\bibinfo {author} {\bibfnamefont {B.~P.}\ \bibnamefont
  {Abbott}} \emph {et~al.} (\bibinfo {collaboration} {Virgo, LIGO
  Scientific}),\ }\href {\doibase 10.3847/2041-8213/aa9f0c} {\bibfield
  {journal} {\bibinfo  {journal} {Astrophys. J.}\ }\textbf {\bibinfo {volume}
  {851}},\ \bibinfo {pages} {L35} (\bibinfo {year} {2017}{\natexlab{b}})},\
  \Eprint {http://arxiv.org/abs/1711.05578} {arXiv:1711.05578 [astro-ph.HE]}
  \BibitemShut {NoStop}%
\bibitem [{\citenamefont {Abbott}\ \emph
  {et~al.}(2017{\natexlab{c}})\citenamefont {Abbott} \emph
  {et~al.}}]{Abbott:2017oio}%
  \BibitemOpen
  \bibfield  {author} {\bibinfo {author} {\bibfnamefont {B.~P.}\ \bibnamefont
  {Abbott}} \emph {et~al.} (\bibinfo {collaboration} {Virgo, LIGO
  Scientific}),\ }\href {\doibase 10.1103/PhysRevLett.119.141101} {\bibfield
  {journal} {\bibinfo  {journal} {Phys. Rev. Lett.}\ }\textbf {\bibinfo
  {volume} {119}},\ \bibinfo {pages} {141101} (\bibinfo {year}
  {2017}{\natexlab{c}})},\ \Eprint {http://arxiv.org/abs/1709.09660}
  {arXiv:1709.09660 [gr-qc]} \BibitemShut {NoStop}%
\bibitem [{\citenamefont {Abbott}\ \emph
  {et~al.}(2017{\natexlab{d}})\citenamefont {Abbott} \emph
  {et~al.}}]{TheLIGOScientific:2017qsa}%
  \BibitemOpen
  \bibfield  {author} {\bibinfo {author} {\bibfnamefont {B.}~\bibnamefont
  {Abbott}} \emph {et~al.} (\bibinfo {collaboration} {Virgo, LIGO
  Scientific}),\ }\href {\doibase 10.1103/PhysRevLett.119.161101} {\bibfield
  {journal} {\bibinfo  {journal} {Phys. Rev. Lett.}\ }\textbf {\bibinfo
  {volume} {119}},\ \bibinfo {pages} {161101} (\bibinfo {year}
  {2017}{\natexlab{d}})},\ \Eprint {http://arxiv.org/abs/1710.05832}
  {arXiv:1710.05832 [gr-qc]} \BibitemShut {NoStop}%
\bibitem [{\citenamefont {Abbott}\ \emph
  {et~al.}(2017{\natexlab{e}})\citenamefont {Abbott} \emph
  {et~al.}}]{GBM:2017lvd}%
  \BibitemOpen
  \bibfield  {author} {\bibinfo {author} {\bibfnamefont {B.~P.}\ \bibnamefont
  {Abbott}} \emph {et~al.} (\bibinfo {collaboration} {GROND, SALT Group,
  OzGrav, DFN, DES, INTEGRAL, Virgo, Insight-Hxmt, MAXI Team, Fermi-LAT, J-GEM,
  RATIR, IceCube, CAASTRO, LWA, ePESSTO, GRAWITA, RIMAS, SKA South
  Africa/MeerKAT, H.E.S.S., 1M2H Team, IKI-GW Follow-up, Fermi GBM, Pi of Sky,
  DWF (Deeper Wider Faster Program), MASTER, AstroSat Cadmium Zinc Telluride
  Imager Team, Swift, Pierre Auger, ASKAP, VINROUGE, JAGWAR, Chandra Team at
  McGill University, TTU-NRAO, GROWTH, AGILE Team, MWA, ATCA, AST3, TOROS,
  Pan-STARRS, NuSTAR, ATLAS Telescopes, BOOTES, CaltechNRAO, LIGO Scientific,
  High Time Resolution Universe Survey, Nordic Optical Telescope, Las Cumbres
  Observatory Group, TZAC Consortium, LOFAR, IPN, DLT40, Texas Tech University,
  HAWC, ANTARES, KU, Dark Energy Camera GW-EM, CALET, Euro VLBI Team, ALMA}),\
  }\href {\doibase 10.3847/2041-8213/aa91c9} {\bibfield  {journal} {\bibinfo
  {journal} {Astrophys. J.}\ }\textbf {\bibinfo {volume} {848}},\ \bibinfo
  {pages} {L12} (\bibinfo {year} {2017}{\natexlab{e}})},\ \Eprint
  {http://arxiv.org/abs/1710.05833} {arXiv:1710.05833 [astro-ph.HE]}
  \BibitemShut {NoStop}%
\bibitem [{\citenamefont {Matzner}(1968)}]{Matzner:1968}%
  \BibitemOpen
  \bibfield  {author} {\bibinfo {author} {\bibfnamefont {R.~A.}\ \bibnamefont
  {Matzner}},\ }\href@noop {} {\bibfield  {journal} {\bibinfo  {journal}
  {Journal of Mathematical Physics}\ }\textbf {\bibinfo {volume} {9}},\
  \bibinfo {pages} {163} (\bibinfo {year} {1968})}\BibitemShut {NoStop}%
\bibitem [{\citenamefont {Sanchez}(1978)}]{Sanchez:1977vz}%
  \BibitemOpen
  \bibfield  {author} {\bibinfo {author} {\bibfnamefont {N.~G.}\ \bibnamefont
  {Sanchez}},\ }\href {\doibase 10.1103/PhysRevD.18.1798} {\bibfield  {journal}
  {\bibinfo  {journal} {Phys. Rev.}\ }\textbf {\bibinfo {volume} {D18}},\
  \bibinfo {pages} {1798} (\bibinfo {year} {1978})}\BibitemShut {NoStop}%
\bibitem [{\citenamefont {Handler}\ and\ \citenamefont
  {Matzner}(1980)}]{Handler:1980un}%
  \BibitemOpen
  \bibfield  {author} {\bibinfo {author} {\bibfnamefont {F.~A.}\ \bibnamefont
  {Handler}}\ and\ \bibinfo {author} {\bibfnamefont {R.~A.}\ \bibnamefont
  {Matzner}},\ }\href {\doibase 10.1103/PhysRevD.22.2331} {\bibfield  {journal}
  {\bibinfo  {journal} {Phys. Rev.}\ }\textbf {\bibinfo {volume} {D22}},\
  \bibinfo {pages} {2331} (\bibinfo {year} {1980})}\BibitemShut {NoStop}%
\bibitem [{\citenamefont {Zhang}\ and\ \citenamefont
  {DeWitt-Morette}(1984)}]{Zhang:1984vt}%
  \BibitemOpen
  \bibfield  {author} {\bibinfo {author} {\bibfnamefont {T.~R.}\ \bibnamefont
  {Zhang}}\ and\ \bibinfo {author} {\bibfnamefont {C.}~\bibnamefont
  {DeWitt-Morette}},\ }\href {\doibase 10.1103/PhysRevLett.52.2313} {\bibfield
  {journal} {\bibinfo  {journal} {Phys. Rev. Lett.}\ }\textbf {\bibinfo
  {volume} {52}},\ \bibinfo {pages} {2313} (\bibinfo {year}
  {1984})}\BibitemShut {NoStop}%
\bibitem [{\citenamefont {Matzner}\ \emph {et~al.}(1985)\citenamefont
  {Matzner}, \citenamefont {DeWitt-Morette}, \citenamefont {Nelson},\ and\
  \citenamefont {Zhang}}]{Matzner:1985rjn}%
  \BibitemOpen
  \bibfield  {author} {\bibinfo {author} {\bibfnamefont {R.~A.}\ \bibnamefont
  {Matzner}}, \bibinfo {author} {\bibfnamefont {C.}~\bibnamefont
  {DeWitt-Morette}}, \bibinfo {author} {\bibfnamefont {B.}~\bibnamefont
  {Nelson}}, \ and\ \bibinfo {author} {\bibfnamefont {T.-R.}\ \bibnamefont
  {Zhang}},\ }\href {\doibase 10.1103/PhysRevD.31.1869} {\bibfield  {journal}
  {\bibinfo  {journal} {Phys. Rev.}\ }\textbf {\bibinfo {volume} {D31}},\
  \bibinfo {pages} {1869} (\bibinfo {year} {1985})}\BibitemShut {NoStop}%
\bibitem [{\citenamefont {Futterman}\ \emph {et~al.}(2012)\citenamefont
  {Futterman}, \citenamefont {Handler},\ and\ \citenamefont
  {Matzner}}]{Futterman:1988ni}%
  \BibitemOpen
  \bibfield  {author} {\bibinfo {author} {\bibfnamefont {J.~A.~H.}\
  \bibnamefont {Futterman}}, \bibinfo {author} {\bibfnamefont {F.~A.}\
  \bibnamefont {Handler}}, \ and\ \bibinfo {author} {\bibfnamefont {R.~A.}\
  \bibnamefont {Matzner}},\ }\href@noop {} {\emph {\bibinfo {title}
  {{Scattering from black holes}}}}\ (\bibinfo  {publisher} {Cambridge
  University Press},\ \bibinfo {year} {2012})\BibitemShut {NoStop}%
\bibitem [{\citenamefont {Crispino}\ \emph
  {et~al.}(2009{\natexlab{a}})\citenamefont {Crispino}, \citenamefont {Dolan},\
  and\ \citenamefont {Oliveira}}]{Crispino:2009ki}%
  \BibitemOpen
  \bibfield  {author} {\bibinfo {author} {\bibfnamefont {L.~C.~B.}\
  \bibnamefont {Crispino}}, \bibinfo {author} {\bibfnamefont {S.~R.}\
  \bibnamefont {Dolan}}, \ and\ \bibinfo {author} {\bibfnamefont {E.~S.}\
  \bibnamefont {Oliveira}},\ }\href {\doibase 10.1103/PhysRevD.79.064022}
  {\bibfield  {journal} {\bibinfo  {journal} {Phys. Rev.}\ }\textbf {\bibinfo
  {volume} {D79}},\ \bibinfo {pages} {064022} (\bibinfo {year}
  {2009}{\natexlab{a}})},\ \Eprint {http://arxiv.org/abs/0904.0999}
  {arXiv:0904.0999 [gr-qc]} \BibitemShut {NoStop}%
\bibitem [{\citenamefont {Crispino}\ \emph
  {et~al.}(2009{\natexlab{b}})\citenamefont {Crispino}, \citenamefont {Dolan},\
  and\ \citenamefont {Oliveira}}]{Crispino:2009xt}%
  \BibitemOpen
  \bibfield  {author} {\bibinfo {author} {\bibfnamefont {L.~C.~B.}\
  \bibnamefont {Crispino}}, \bibinfo {author} {\bibfnamefont {S.~R.}\
  \bibnamefont {Dolan}}, \ and\ \bibinfo {author} {\bibfnamefont {E.~S.}\
  \bibnamefont {Oliveira}},\ }\href {\doibase 10.1103/PhysRevLett.102.231103}
  {\bibfield  {journal} {\bibinfo  {journal} {Phys. Rev. Lett.}\ }\textbf
  {\bibinfo {volume} {102}},\ \bibinfo {pages} {231103} (\bibinfo {year}
  {2009}{\natexlab{b}})},\ \Eprint {http://arxiv.org/abs/0905.3339}
  {arXiv:0905.3339 [gr-qc]} \BibitemShut {NoStop}%
\bibitem [{\citenamefont {Crispino}\ \emph {et~al.}(2014)\citenamefont
  {Crispino}, \citenamefont {Dolan}, \citenamefont {Higuchi},\ and\
  \citenamefont {de~Oliveira}}]{Crispino:2014eea}%
  \BibitemOpen
  \bibfield  {author} {\bibinfo {author} {\bibfnamefont {L.~C.~B.}\
  \bibnamefont {Crispino}}, \bibinfo {author} {\bibfnamefont {S.~R.}\
  \bibnamefont {Dolan}}, \bibinfo {author} {\bibfnamefont {A.}~\bibnamefont
  {Higuchi}}, \ and\ \bibinfo {author} {\bibfnamefont {E.~S.}\ \bibnamefont
  {de~Oliveira}},\ }\href {\doibase 10.1103/PhysRevD.90.064027} {\bibfield
  {journal} {\bibinfo  {journal} {Phys. Rev.}\ }\textbf {\bibinfo {volume}
  {D90}},\ \bibinfo {pages} {064027} (\bibinfo {year} {2014})},\ \Eprint
  {http://arxiv.org/abs/1409.4803} {arXiv:1409.4803 [gr-qc]} \BibitemShut
  {NoStop}%
\bibitem [{\citenamefont {Crispino}\ \emph {et~al.}(2015)\citenamefont
  {Crispino}, \citenamefont {Dolan}, \citenamefont {Higuchi},\ and\
  \citenamefont {de~Oliveira}}]{Crispino:2015gua}%
  \BibitemOpen
  \bibfield  {author} {\bibinfo {author} {\bibfnamefont {L.~C.~B.}\
  \bibnamefont {Crispino}}, \bibinfo {author} {\bibfnamefont {S.~R.}\
  \bibnamefont {Dolan}}, \bibinfo {author} {\bibfnamefont {A.}~\bibnamefont
  {Higuchi}}, \ and\ \bibinfo {author} {\bibfnamefont {E.~S.}\ \bibnamefont
  {de~Oliveira}},\ }\href {\doibase 10.1103/PhysRevD.92.084056} {\bibfield
  {journal} {\bibinfo  {journal} {Phys. Rev.}\ }\textbf {\bibinfo {volume}
  {D92}},\ \bibinfo {pages} {084056} (\bibinfo {year} {2015})},\ \Eprint
  {http://arxiv.org/abs/1507.03993} {arXiv:1507.03993 [gr-qc]} \BibitemShut
  {NoStop}%
\bibitem [{\citenamefont {Leite}\ \emph {et~al.}(2017)\citenamefont {Leite},
  \citenamefont {Dolan},\ and\ \citenamefont {Crispino}}]{Leite:2017zyb}%
  \BibitemOpen
  \bibfield  {author} {\bibinfo {author} {\bibfnamefont {L.~C.~S.}\
  \bibnamefont {Leite}}, \bibinfo {author} {\bibfnamefont {S.~R.}\ \bibnamefont
  {Dolan}}, \ and\ \bibinfo {author} {\bibfnamefont {L.~C.~B.}\ \bibnamefont
  {Crispino}},\ }\href {\doibase 10.1016/j.physletb.2017.09.048} {\bibfield
  {journal} {\bibinfo  {journal} {Phys. Lett.}\ }\textbf {\bibinfo {volume}
  {B774}},\ \bibinfo {pages} {130} (\bibinfo {year} {2017})},\ \Eprint
  {http://arxiv.org/abs/1707.01144} {arXiv:1707.01144 [gr-qc]} \BibitemShut
  {NoStop}%
\bibitem [{\citenamefont {Dolan}(2008{\natexlab{a}})}]{Dolan:2007ut}%
  \BibitemOpen
  \bibfield  {author} {\bibinfo {author} {\bibfnamefont {S.~R.}\ \bibnamefont
  {Dolan}},\ }\href {\doibase 10.1103/PhysRevD.77.044004} {\bibfield  {journal}
  {\bibinfo  {journal} {Phys. Rev.}\ }\textbf {\bibinfo {volume} {D77}},\
  \bibinfo {pages} {044004} (\bibinfo {year} {2008}{\natexlab{a}})},\ \Eprint
  {http://arxiv.org/abs/0710.4252} {arXiv:0710.4252 [gr-qc]} \BibitemShut
  {NoStop}%
\bibitem [{\citenamefont {Dolan}(2008{\natexlab{b}})}]{Dolan:2008kf}%
  \BibitemOpen
  \bibfield  {author} {\bibinfo {author} {\bibfnamefont {S.~R.}\ \bibnamefont
  {Dolan}},\ }\href {\doibase 10.1088/0264-9381/25/23/235002} {\bibfield
  {journal} {\bibinfo  {journal} {Class. Quant. Grav.}\ }\textbf {\bibinfo
  {volume} {25}},\ \bibinfo {pages} {235002} (\bibinfo {year}
  {2008}{\natexlab{b}})},\ \Eprint {http://arxiv.org/abs/0801.3805}
  {arXiv:0801.3805 [gr-qc]} \BibitemShut {NoStop}%
\bibitem [{\citenamefont {Kanai}\ and\ \citenamefont
  {Nambu}(2013)}]{Kanai:2013rga}%
  \BibitemOpen
  \bibfield  {author} {\bibinfo {author} {\bibfnamefont {K.-i.}\ \bibnamefont
  {Kanai}}\ and\ \bibinfo {author} {\bibfnamefont {Y.}~\bibnamefont {Nambu}},\
  }\href {\doibase 10.1088/0264-9381/30/17/175002} {\bibfield  {journal}
  {\bibinfo  {journal} {Class. Quant. Grav.}\ }\textbf {\bibinfo {volume}
  {30}},\ \bibinfo {pages} {175002} (\bibinfo {year} {2013})},\ \Eprint
  {http://arxiv.org/abs/1303.5520} {arXiv:1303.5520 [gr-qc]} \BibitemShut
  {NoStop}%
\bibitem [{\citenamefont {Nambu}\ and\ \citenamefont
  {Noda}(2016)}]{Nambu:2015aea}%
  \BibitemOpen
  \bibfield  {author} {\bibinfo {author} {\bibfnamefont {Y.}~\bibnamefont
  {Nambu}}\ and\ \bibinfo {author} {\bibfnamefont {S.}~\bibnamefont {Noda}},\
  }\href {\doibase 10.1088/0264-9381/33/7/075011} {\bibfield  {journal}
  {\bibinfo  {journal} {Class. Quant. Grav.}\ }\textbf {\bibinfo {volume}
  {33}},\ \bibinfo {pages} {075011} (\bibinfo {year} {2016})},\ \Eprint
  {http://arxiv.org/abs/1502.05468} {arXiv:1502.05468 [gr-qc]} \BibitemShut
  {NoStop}%
\bibitem [{\citenamefont {Rosa}(2017)}]{Rosa:2016bli}%
  \BibitemOpen
  \bibfield  {author} {\bibinfo {author} {\bibfnamefont {J.~G.}\ \bibnamefont
  {Rosa}},\ }\href {\doibase 10.1103/PhysRevD.95.064017} {\bibfield  {journal}
  {\bibinfo  {journal} {Phys. Rev.}\ }\textbf {\bibinfo {volume} {D95}},\
  \bibinfo {pages} {064017} (\bibinfo {year} {2017})},\ \Eprint
  {http://arxiv.org/abs/1612.01826} {arXiv:1612.01826 [gr-qc]} \BibitemShut
  {NoStop}%
\bibitem [{\citenamefont {Alexandre}\ and\ \citenamefont
  {Clough}(2018)}]{Alexandre:2018crg}%
  \BibitemOpen
  \bibfield  {author} {\bibinfo {author} {\bibfnamefont {J.}~\bibnamefont
  {Alexandre}}\ and\ \bibinfo {author} {\bibfnamefont {K.}~\bibnamefont
  {Clough}},\ }\href {\doibase 10.1103/PhysRevD.98.043004} {\bibfield
  {journal} {\bibinfo  {journal} {Phys. Rev.}\ }\textbf {\bibinfo {volume}
  {D98}},\ \bibinfo {pages} {043004} (\bibinfo {year} {2018})},\ \Eprint
  {http://arxiv.org/abs/1805.01874} {arXiv:1805.01874 [hep-ph]} \BibitemShut
  {NoStop}%
\bibitem [{\citenamefont {Sporea}(2018)}]{Sporea:2018rsk}%
  \BibitemOpen
  \bibfield  {author} {\bibinfo {author} {\bibfnamefont {C.~A.}\ \bibnamefont
  {Sporea}},\ }\href@noop {} {\  (\bibinfo {year} {2018})},\ \Eprint
  {http://arxiv.org/abs/1812.09945} {arXiv:1812.09945 [gr-qc]} \BibitemShut
  {NoStop}%
\bibitem [{\citenamefont {Leite}\ \emph {et~al.}(2019)\citenamefont {Leite},
  \citenamefont {Macedo},\ and\ \citenamefont {Crispino}}]{Leite:2019uql}%
  \BibitemOpen
  \bibfield  {author} {\bibinfo {author} {\bibfnamefont {L.~C.~S.}\
  \bibnamefont {Leite}}, \bibinfo {author} {\bibfnamefont {C.~F.~B.}\
  \bibnamefont {Macedo}}, \ and\ \bibinfo {author} {\bibfnamefont {L.~C.~B.}\
  \bibnamefont {Crispino}},\ }\href@noop {} {\  (\bibinfo {year} {2019})},\
  \Eprint {http://arxiv.org/abs/1901.07074} {arXiv:1901.07074 [gr-qc]}
  \BibitemShut {NoStop}%
\bibitem [{\citenamefont {Dolan}\ and\ \citenamefont
  {Stratton}(2017)}]{Dolan:2017rtj}%
  \BibitemOpen
  \bibfield  {author} {\bibinfo {author} {\bibfnamefont {S.~R.}\ \bibnamefont
  {Dolan}}\ and\ \bibinfo {author} {\bibfnamefont {T.}~\bibnamefont
  {Stratton}},\ }\href {\doibase 10.1103/PhysRevD.95.124055} {\bibfield
  {journal} {\bibinfo  {journal} {Phys. Rev.}\ }\textbf {\bibinfo {volume}
  {D95}},\ \bibinfo {pages} {124055} (\bibinfo {year} {2017})},\ \Eprint
  {http://arxiv.org/abs/1702.06127} {arXiv:1702.06127 [gr-qc]} \BibitemShut
  {NoStop}%
\bibitem [{\citenamefont {Cotaescu}\ and\ \citenamefont
  {Sporea}(2019)}]{Cotaescu:2018etx}%
  \BibitemOpen
  \bibfield  {author} {\bibinfo {author} {\bibfnamefont {I.~I.}\ \bibnamefont
  {Cotaescu}}\ and\ \bibinfo {author} {\bibfnamefont {C.~A.}\ \bibnamefont
  {Sporea}},\ }\href {\doibase 10.1140/epjc/s10052-018-6525-2} {\bibfield
  {journal} {\bibinfo  {journal} {Eur. Phys. J.}\ }\textbf {\bibinfo {volume}
  {C79}},\ \bibinfo {pages} {15} (\bibinfo {year} {2019})},\ \Eprint
  {http://arxiv.org/abs/1811.07723} {arXiv:1811.07723 [gr-qc]} \BibitemShut
  {NoStop}%
\bibitem [{\citenamefont {Tominaga}\ \emph {et~al.}(1999)\citenamefont
  {Tominaga}, \citenamefont {Saijo},\ and\ \citenamefont
  {Maeda}}]{Tominaga:1999iy}%
  \BibitemOpen
  \bibfield  {author} {\bibinfo {author} {\bibfnamefont {K.}~\bibnamefont
  {Tominaga}}, \bibinfo {author} {\bibfnamefont {M.}~\bibnamefont {Saijo}}, \
  and\ \bibinfo {author} {\bibfnamefont {K.-i.}\ \bibnamefont {Maeda}},\ }\href
  {\doibase 10.1103/PhysRevD.60.024004} {\bibfield  {journal} {\bibinfo
  {journal} {Phys. Rev.}\ }\textbf {\bibinfo {volume} {D60}},\ \bibinfo {pages}
  {024004} (\bibinfo {year} {1999})},\ \Eprint
  {http://arxiv.org/abs/gr-qc/9901040} {arXiv:gr-qc/9901040 [gr-qc]}
  \BibitemShut {NoStop}%
\bibitem [{\citenamefont {Tominaga}\ \emph {et~al.}(2001)\citenamefont
  {Tominaga}, \citenamefont {Saijo},\ and\ \citenamefont
  {Maeda}}]{Tominaga:2000cs}%
  \BibitemOpen
  \bibfield  {author} {\bibinfo {author} {\bibfnamefont {K.}~\bibnamefont
  {Tominaga}}, \bibinfo {author} {\bibfnamefont {M.}~\bibnamefont {Saijo}}, \
  and\ \bibinfo {author} {\bibfnamefont {K.-i.}\ \bibnamefont {Maeda}},\ }\href
  {\doibase 10.1103/PhysRevD.63.124012} {\bibfield  {journal} {\bibinfo
  {journal} {Phys. Rev.}\ }\textbf {\bibinfo {volume} {D63}},\ \bibinfo {pages}
  {124012} (\bibinfo {year} {2001})},\ \Eprint
  {http://arxiv.org/abs/gr-qc/0009055} {arXiv:gr-qc/0009055 [gr-qc]}
  \BibitemShut {NoStop}%
\bibitem [{\citenamefont {Bernuzzi}\ \emph {et~al.}(2008)\citenamefont
  {Bernuzzi}, \citenamefont {Nagar},\ and\ \citenamefont
  {De~Pietri}}]{Bernuzzi:2008rq}%
  \BibitemOpen
  \bibfield  {author} {\bibinfo {author} {\bibfnamefont {S.}~\bibnamefont
  {Bernuzzi}}, \bibinfo {author} {\bibfnamefont {A.}~\bibnamefont {Nagar}}, \
  and\ \bibinfo {author} {\bibfnamefont {R.}~\bibnamefont {De~Pietri}},\ }\href
  {\doibase 10.1103/PhysRevD.77.044042} {\bibfield  {journal} {\bibinfo
  {journal} {Phys. Rev.}\ }\textbf {\bibinfo {volume} {D77}},\ \bibinfo {pages}
  {044042} (\bibinfo {year} {2008})},\ \Eprint {http://arxiv.org/abs/0801.2090}
  {arXiv:0801.2090 [gr-qc]} \BibitemShut {NoStop}%
\bibitem [{\citenamefont {Ipser}\ and\ \citenamefont
  {Price}(1991)}]{Ipser:1991ind}%
  \BibitemOpen
  \bibfield  {author} {\bibinfo {author} {\bibfnamefont {J.~R.}\ \bibnamefont
  {Ipser}}\ and\ \bibinfo {author} {\bibfnamefont {R.~H.}\ \bibnamefont
  {Price}},\ }\href {\doibase 10.1103/PhysRevD.43.1768} {\bibfield  {journal}
  {\bibinfo  {journal} {Phys. Rev.}\ }\textbf {\bibinfo {volume} {D43}},\
  \bibinfo {pages} {1768} (\bibinfo {year} {1991})}\BibitemShut {NoStop}%
\bibitem [{\citenamefont {Kojima}(1992)}]{Kojima:1992ie}%
  \BibitemOpen
  \bibfield  {author} {\bibinfo {author} {\bibfnamefont {Y.}~\bibnamefont
  {Kojima}},\ }\href {\doibase 10.1103/PhysRevD.46.4289} {\bibfield  {journal}
  {\bibinfo  {journal} {Phys. Rev.}\ }\textbf {\bibinfo {volume} {D46}},\
  \bibinfo {pages} {4289} (\bibinfo {year} {1992})}\BibitemShut {NoStop}%
\bibitem [{\citenamefont {Allen}\ \emph {et~al.}(1998)\citenamefont {Allen},
  \citenamefont {Andersson}, \citenamefont {Kokkotas},\ and\ \citenamefont
  {Schutz}}]{Allen:1997xj}%
  \BibitemOpen
  \bibfield  {author} {\bibinfo {author} {\bibfnamefont {G.}~\bibnamefont
  {Allen}}, \bibinfo {author} {\bibfnamefont {N.}~\bibnamefont {Andersson}},
  \bibinfo {author} {\bibfnamefont {K.~D.}\ \bibnamefont {Kokkotas}}, \ and\
  \bibinfo {author} {\bibfnamefont {B.~F.}\ \bibnamefont {Schutz}},\ }\href
  {\doibase 10.1103/PhysRevD.58.124012} {\bibfield  {journal} {\bibinfo
  {journal} {Phys. Rev.}\ }\textbf {\bibinfo {volume} {D58}},\ \bibinfo {pages}
  {124012} (\bibinfo {year} {1998})},\ \Eprint
  {http://arxiv.org/abs/gr-qc/9704023} {arXiv:gr-qc/9704023 [gr-qc]}
  \BibitemShut {NoStop}%
\bibitem [{\citenamefont {Martel}\ and\ \citenamefont
  {Poisson}(2005)}]{Martel:2005ir}%
  \BibitemOpen
  \bibfield  {author} {\bibinfo {author} {\bibfnamefont {K.}~\bibnamefont
  {Martel}}\ and\ \bibinfo {author} {\bibfnamefont {E.}~\bibnamefont
  {Poisson}},\ }\href {\doibase 10.1103/PhysRevD.71.104003} {\bibfield
  {journal} {\bibinfo  {journal} {Phys. Rev.}\ }\textbf {\bibinfo {volume}
  {D71}},\ \bibinfo {pages} {104003} (\bibinfo {year} {2005})},\ \Eprint
  {http://arxiv.org/abs/gr-qc/0502028} {arXiv:gr-qc/0502028 [gr-qc]}
  \BibitemShut {NoStop}%
\bibitem [{\citenamefont {Barack}\ and\ \citenamefont
  {Lousto}(2005)}]{Barack:2005nr}%
  \BibitemOpen
  \bibfield  {author} {\bibinfo {author} {\bibfnamefont {L.}~\bibnamefont
  {Barack}}\ and\ \bibinfo {author} {\bibfnamefont {C.~O.}\ \bibnamefont
  {Lousto}},\ }\href {\doibase 10.1103/PhysRevD.72.104026} {\bibfield
  {journal} {\bibinfo  {journal} {Phys. Rev.}\ }\textbf {\bibinfo {volume}
  {D72}},\ \bibinfo {pages} {104026} (\bibinfo {year} {2005})},\ \Eprint
  {http://arxiv.org/abs/gr-qc/0510019} {arXiv:gr-qc/0510019 [gr-qc]}
  \BibitemShut {NoStop}%
\bibitem [{\citenamefont {Cunningham}\ \emph {et~al.}(1978)\citenamefont
  {Cunningham}, \citenamefont {Price},\ and\ \citenamefont
  {Moncrief}}]{Cunningham:1978zfa}%
  \BibitemOpen
  \bibfield  {author} {\bibinfo {author} {\bibfnamefont {C.~T.}\ \bibnamefont
  {Cunningham}}, \bibinfo {author} {\bibfnamefont {R.~H.}\ \bibnamefont
  {Price}}, \ and\ \bibinfo {author} {\bibfnamefont {V.}~\bibnamefont
  {Moncrief}},\ }\href {\doibase 10.1086/156413} {\bibfield  {journal}
  {\bibinfo  {journal} {Astrophys. J.}\ }\textbf {\bibinfo {volume} {224}},\
  \bibinfo {pages} {643} (\bibinfo {year} {1978})}\BibitemShut {NoStop}%
\bibitem [{\citenamefont {Cunningham}\ \emph {et~al.}(1979)\citenamefont
  {Cunningham}, \citenamefont {Price},\ and\ \citenamefont
  {Moncrief}}]{Cunningham:1979px}%
  \BibitemOpen
  \bibfield  {author} {\bibinfo {author} {\bibfnamefont {C.~T.}\ \bibnamefont
  {Cunningham}}, \bibinfo {author} {\bibfnamefont {R.~H.}\ \bibnamefont
  {Price}}, \ and\ \bibinfo {author} {\bibfnamefont {V.}~\bibnamefont
  {Moncrief}},\ }\href {\doibase 10.1086/157147} {\bibfield  {journal}
  {\bibinfo  {journal} {Astrophys. J.}\ }\textbf {\bibinfo {volume} {230}},\
  \bibinfo {pages} {870} (\bibinfo {year} {1979})}\BibitemShut {NoStop}%
\bibitem [{\citenamefont {Chandrasekhar}\ and\ \citenamefont
  {Ferrari}(1991{\natexlab{a}})}]{Chandrasekhar247}%
  \BibitemOpen
  \bibfield  {author} {\bibinfo {author} {\bibfnamefont {S.}~\bibnamefont
  {Chandrasekhar}}\ and\ \bibinfo {author} {\bibfnamefont {V.}~\bibnamefont
  {Ferrari}},\ }\href {\doibase 10.1098/rspa.1991.0016} {\bibfield  {journal}
  {\bibinfo  {journal} {Proceedings of the Royal Society of London A:
  Mathematical, Physical and Engineering Sciences}\ }\textbf {\bibinfo {volume}
  {432}},\ \bibinfo {pages} {247} (\bibinfo {year} {1991}{\natexlab{a}})},\
  \Eprint
  {http://arxiv.org/abs/http://rspa.royalsocietypublishing.org/content/432/1885/247.full.pdf}
  {http://rspa.royalsocietypublishing.org/content/432/1885/247.full.pdf}
  \BibitemShut {NoStop}%
\bibitem [{\citenamefont {Chandrasekhar}\ and\ \citenamefont
  {Ferrari}(1991{\natexlab{b}})}]{Chandrasekhar247b}%
  \BibitemOpen
  \bibfield  {author} {\bibinfo {author} {\bibfnamefont {S.}~\bibnamefont
  {Chandrasekhar}}\ and\ \bibinfo {author} {\bibfnamefont {V.}~\bibnamefont
  {Ferrari}},\ }\href {\doibase 10.1098/rspa.1991.0016} {\bibfield  {journal}
  {\bibinfo  {journal} {Proc. Roy. Soc. London Ser.A}\ }\textbf {\bibinfo
  {volume} {432}},\ \bibinfo {pages} {247} (\bibinfo {year}
  {1991}{\natexlab{b}})}\BibitemShut {NoStop}%
\bibitem [{\citenamefont {Chandrasekhar}\ and\ \citenamefont
  {Ferrari}(1991{\natexlab{c}})}]{Chandrasekhar449}%
  \BibitemOpen
  \bibfield  {author} {\bibinfo {author} {\bibfnamefont {S.}~\bibnamefont
  {Chandrasekhar}}\ and\ \bibinfo {author} {\bibfnamefont {V.}~\bibnamefont
  {Ferrari}},\ }\href {\doibase 10.1098/rspa.1991.0104} {\bibfield  {journal}
  {\bibinfo  {journal} {Proceedings of the Royal Society of London A:
  Mathematical, Physical and Engineering Sciences}\ }\textbf {\bibinfo {volume}
  {434}},\ \bibinfo {pages} {449} (\bibinfo {year} {1991}{\natexlab{c}})},\
  \Eprint
  {http://arxiv.org/abs/http://rspa.royalsocietypublishing.org/content/434/1891/449.full.pdf}
  {http://rspa.royalsocietypublishing.org/content/434/1891/449.full.pdf}
  \BibitemShut {NoStop}%
\bibitem [{\citenamefont {Halder}\ \emph {et~al.}(2019)\citenamefont {Halder},
  \citenamefont {Banerjee},\ and\ \citenamefont {Majumdar}}]{Halder:2019cmp}%
  \BibitemOpen
  \bibfield  {author} {\bibinfo {author} {\bibfnamefont {A.}~\bibnamefont
  {Halder}}, \bibinfo {author} {\bibfnamefont {S.}~\bibnamefont {Banerjee}}, \
  and\ \bibinfo {author} {\bibfnamefont {D.}~\bibnamefont {Majumdar}},\
  }\href@noop {} {\  (\bibinfo {year} {2019})},\ \Eprint
  {http://arxiv.org/abs/1902.06903} {arXiv:1902.06903 [gr-qc]} \BibitemShut
  {NoStop}%
\bibitem [{\citenamefont {Airy}\ \emph {et~al.}(1838)\citenamefont {Airy} \emph
  {et~al.}}]{Airy:1838}%
  \BibitemOpen
  \bibfield  {author} {\bibinfo {author} {\bibfnamefont {G.~B.}\ \bibnamefont
  {Airy}} \emph {et~al.},\ }\href@noop {} {\bibfield  {journal} {\bibinfo
  {journal} {Transactions of the Cambridge Philosophical Society}\ }\textbf
  {\bibinfo {volume} {6}},\ \bibinfo {pages} {379} (\bibinfo {year}
  {1838})}\BibitemShut {NoStop}%
\bibitem [{\citenamefont {Ford}\ and\ \citenamefont
  {Wheeler}(1959)}]{Ford:1959}%
  \BibitemOpen
  \bibfield  {author} {\bibinfo {author} {\bibfnamefont {K.~W.}\ \bibnamefont
  {Ford}}\ and\ \bibinfo {author} {\bibfnamefont {J.~A.}\ \bibnamefont
  {Wheeler}},\ }\href@noop {} {\bibfield  {journal} {\bibinfo  {journal}
  {Annals of Physics}\ }\textbf {\bibinfo {volume} {7}},\ \bibinfo {pages}
  {259} (\bibinfo {year} {1959})}\BibitemShut {NoStop}%
\bibitem [{\citenamefont {Westervelt}(1971)}]{Westervelt:1971pm}%
  \BibitemOpen
  \bibfield  {author} {\bibinfo {author} {\bibfnamefont {P.~J.}\ \bibnamefont
  {Westervelt}},\ }\href {\doibase 10.1103/PhysRevD.3.2319} {\bibfield
  {journal} {\bibinfo  {journal} {Phys. Rev.}\ }\textbf {\bibinfo {volume}
  {D3}},\ \bibinfo {pages} {2319} (\bibinfo {year} {1971})}\BibitemShut
  {NoStop}%
\bibitem [{\citenamefont {Peters}(1976)}]{Peters:1976jx}%
  \BibitemOpen
  \bibfield  {author} {\bibinfo {author} {\bibfnamefont {P.~C.}\ \bibnamefont
  {Peters}},\ }\href {\doibase 10.1103/PhysRevD.13.775} {\bibfield  {journal}
  {\bibinfo  {journal} {Phys. Rev.}\ }\textbf {\bibinfo {volume} {D13}},\
  \bibinfo {pages} {775} (\bibinfo {year} {1976})}\BibitemShut {NoStop}%
\bibitem [{\citenamefont {De~Logi}\ and\ \citenamefont
  {Kovacs}(1977)}]{DeLogi:1977dp}%
  \BibitemOpen
  \bibfield  {author} {\bibinfo {author} {\bibfnamefont {W.~K.}\ \bibnamefont
  {De~Logi}}\ and\ \bibinfo {author} {\bibfnamefont {S.~J.}\ \bibnamefont
  {Kovacs}},\ }\href {\doibase 10.1103/PhysRevD.16.237} {\bibfield  {journal}
  {\bibinfo  {journal} {Phys. Rev.}\ }\textbf {\bibinfo {volume} {D16}},\
  \bibinfo {pages} {237} (\bibinfo {year} {1977})}\BibitemShut {NoStop}%
\bibitem [{\citenamefont {Regge}\ and\ \citenamefont
  {Wheeler}(1957)}]{Regge:1957td}%
  \BibitemOpen
  \bibfield  {author} {\bibinfo {author} {\bibfnamefont {T.}~\bibnamefont
  {Regge}}\ and\ \bibinfo {author} {\bibfnamefont {J.~A.}\ \bibnamefont
  {Wheeler}},\ }\href {\doibase 10.1103/PhysRev.108.1063} {\bibfield  {journal}
  {\bibinfo  {journal} {Phys. Rev.}\ }\textbf {\bibinfo {volume} {108}},\
  \bibinfo {pages} {1063} (\bibinfo {year} {1957})}\BibitemShut {NoStop}%
\bibitem [{\citenamefont {Zerilli}(1970)}]{Zerilli:1971wd}%
  \BibitemOpen
  \bibfield  {author} {\bibinfo {author} {\bibfnamefont {F.~J.}\ \bibnamefont
  {Zerilli}},\ }\href {\doibase 10.1103/PhysRevD.2.2141} {\bibfield  {journal}
  {\bibinfo  {journal} {Phys. Rev.}\ }\textbf {\bibinfo {volume} {D2}},\
  \bibinfo {pages} {2141} (\bibinfo {year} {1970})}\BibitemShut {NoStop}%
\bibitem [{\citenamefont {Vishveshwara}(1970)}]{vishveshwara1970scattering}%
  \BibitemOpen
  \bibfield  {author} {\bibinfo {author} {\bibfnamefont {C.}~\bibnamefont
  {Vishveshwara}},\ }\href@noop {} {\bibfield  {journal} {\bibinfo  {journal}
  {Nature}\ }\textbf {\bibinfo {volume} {227}},\ \bibinfo {pages} {936}
  (\bibinfo {year} {1970})}\BibitemShut {NoStop}%
\bibitem [{\citenamefont {Schutz}(1985)}]{Schutz:1985jx}%
  \BibitemOpen
  \bibfield  {author} {\bibinfo {author} {\bibfnamefont {B.~F.}\ \bibnamefont
  {Schutz}},\ }\href@noop {} {\emph {\bibinfo {title} {{A first course in
  General Relativity}}}}\ (\bibinfo  {publisher} {Cambridge Univ. Pr.},\
  \bibinfo {address} {Cambridge, UK},\ \bibinfo {year} {1985})\BibitemShut
  {NoStop}%
\bibitem [{\citenamefont {Shapiro}\ and\ \citenamefont
  {Teukolsky}(1983)}]{Shapiro:1983du}%
  \BibitemOpen
  \bibfield  {author} {\bibinfo {author} {\bibfnamefont {S.~L.}\ \bibnamefont
  {Shapiro}}\ and\ \bibinfo {author} {\bibfnamefont {S.~A.}\ \bibnamefont
  {Teukolsky}},\ }\href@noop {} {\emph {\bibinfo {title} {{Black holes, white
  dwarfs, and neutron stars: The physics of compact objects}}}}\ (\bibinfo
  {publisher} {Wiley},\ \bibinfo {address} {New York, USA},\ \bibinfo {year}
  {1983})\BibitemShut {NoStop}%
\bibitem [{\citenamefont {Schwarzschild}(1916)}]{Schwarzschild:1916}%
  \BibitemOpen
  \bibfield  {author} {\bibinfo {author} {\bibfnamefont {K.}~\bibnamefont
  {Schwarzschild}},\ }in\ \href@noop {} {\emph {\bibinfo {booktitle}
  {Sitzungsberichte der K{\"o}niglich Preussischen Akademie der Wissenschaften
  zu Berlin, Phys.-Math. Klasse, 424-434 (1916)}}}\ (\bibinfo {year}
  {1916})\BibitemShut {NoStop}%
\bibitem [{\citenamefont {{Tooper}}(1964)}]{1964Tooper}%
  \BibitemOpen
  \bibfield  {author} {\bibinfo {author} {\bibfnamefont {R.~F.}\ \bibnamefont
  {{Tooper}}},\ }\href {\doibase 10.1086/147939} {\bibfield  {journal}
  {\bibinfo  {journal} {\apj}\ }\textbf {\bibinfo {volume} {140}},\ \bibinfo
  {pages} {434} (\bibinfo {year} {1964})}\BibitemShut {NoStop}%
\bibitem [{\citenamefont {Seidel}\ and\ \citenamefont
  {Moore}(1987)}]{Seidel:1987in}%
  \BibitemOpen
  \bibfield  {author} {\bibinfo {author} {\bibfnamefont {E.}~\bibnamefont
  {Seidel}}\ and\ \bibinfo {author} {\bibfnamefont {T.}~\bibnamefont {Moore}},\
  }\href {\doibase 10.1103/PhysRevD.35.2287} {\bibfield  {journal} {\bibinfo
  {journal} {Phys. Rev.}\ }\textbf {\bibinfo {volume} {D35}},\ \bibinfo {pages}
  {2287} (\bibinfo {year} {1987})}\BibitemShut {NoStop}%
\bibitem [{\citenamefont {Seidel}(1990)}]{Seidel:1990xb}%
  \BibitemOpen
  \bibfield  {author} {\bibinfo {author} {\bibfnamefont {E.}~\bibnamefont
  {Seidel}},\ }\href {\doibase 10.1103/PhysRevD.42.1884} {\bibfield  {journal}
  {\bibinfo  {journal} {Phys. Rev.}\ }\textbf {\bibinfo {volume} {D42}},\
  \bibinfo {pages} {1884} (\bibinfo {year} {1990})}\BibitemShut {NoStop}%
\bibitem [{\citenamefont {Thorne}\ and\ \citenamefont
  {Campolattaro}(1967)}]{thorne1967non}%
  \BibitemOpen
  \bibfield  {author} {\bibinfo {author} {\bibfnamefont {K.~S.}\ \bibnamefont
  {Thorne}}\ and\ \bibinfo {author} {\bibfnamefont {A.}~\bibnamefont
  {Campolattaro}},\ }\href@noop {} {\bibfield  {journal} {\bibinfo  {journal}
  {The Astrophysical Journal}\ }\textbf {\bibinfo {volume} {149}},\ \bibinfo
  {pages} {591} (\bibinfo {year} {1967})}\BibitemShut {NoStop}%
\bibitem [{\citenamefont {Goldberg}\ \emph {et~al.}(1967)\citenamefont
  {Goldberg}, \citenamefont {MacFarlane}, \citenamefont {Newman}, \citenamefont
  {Rohrlich},\ and\ \citenamefont {Sudarshan}}]{Goldberg:1966uu}%
  \BibitemOpen
  \bibfield  {author} {\bibinfo {author} {\bibfnamefont {J.~N.}\ \bibnamefont
  {Goldberg}}, \bibinfo {author} {\bibfnamefont {A.~J.}\ \bibnamefont
  {MacFarlane}}, \bibinfo {author} {\bibfnamefont {E.~T.}\ \bibnamefont
  {Newman}}, \bibinfo {author} {\bibfnamefont {F.}~\bibnamefont {Rohrlich}}, \
  and\ \bibinfo {author} {\bibfnamefont {E.~C.~G.}\ \bibnamefont {Sudarshan}},\
  }\href {\doibase 10.1063/1.1705135} {\bibfield  {journal} {\bibinfo
  {journal} {J. Math. Phys.}\ }\textbf {\bibinfo {volume} {8}},\ \bibinfo
  {pages} {2155} (\bibinfo {year} {1967})}\BibitemShut {NoStop}%
\bibitem [{\citenamefont {Goldberg}\ and\ \citenamefont
  {Smith}(1972)}]{Goldberg:1972zzb}%
  \BibitemOpen
  \bibfield  {author} {\bibinfo {author} {\bibfnamefont {D.~A.}\ \bibnamefont
  {Goldberg}}\ and\ \bibinfo {author} {\bibfnamefont {S.~M.}\ \bibnamefont
  {Smith}},\ }\href {\doibase 10.1103/PhysRevLett.29.500} {\bibfield  {journal}
  {\bibinfo  {journal} {Phys. Rev. Lett.}\ }\textbf {\bibinfo {volume} {29}},\
  \bibinfo {pages} {500} (\bibinfo {year} {1972})}\BibitemShut {NoStop}%
\bibitem [{\citenamefont {Goldberg}\ \emph {et~al.}(1974)\citenamefont
  {Goldberg}, \citenamefont {Smith},\ and\ \citenamefont
  {Burdzik}}]{Goldberg:1974zza}%
  \BibitemOpen
  \bibfield  {author} {\bibinfo {author} {\bibfnamefont {D.~A.}\ \bibnamefont
  {Goldberg}}, \bibinfo {author} {\bibfnamefont {S.~M.}\ \bibnamefont {Smith}},
  \ and\ \bibinfo {author} {\bibfnamefont {G.~F.}\ \bibnamefont {Burdzik}},\
  }\href {\doibase 10.1103/PhysRevC.10.1362} {\bibfield  {journal} {\bibinfo
  {journal} {Phys. Rev.}\ }\textbf {\bibinfo {volume} {C10}},\ \bibinfo {pages}
  {1362} (\bibinfo {year} {1974})}\BibitemShut {NoStop}%
\bibitem [{\citenamefont {Khoa}\ \emph {et~al.}(2007)\citenamefont {Khoa},
  \citenamefont {von Oertzen}, \citenamefont {Bohlen},\ and\ \citenamefont
  {Ohkubo}}]{Khoa:2006id}%
  \BibitemOpen
  \bibfield  {author} {\bibinfo {author} {\bibfnamefont {D.~T.}\ \bibnamefont
  {Khoa}}, \bibinfo {author} {\bibfnamefont {W.}~\bibnamefont {von Oertzen}},
  \bibinfo {author} {\bibfnamefont {H.~G.}\ \bibnamefont {Bohlen}}, \ and\
  \bibinfo {author} {\bibfnamefont {S.}~\bibnamefont {Ohkubo}},\ }\href
  {\doibase 10.1088/0954-3899/34/3/R01} {\bibfield  {journal} {\bibinfo
  {journal} {J. Phys.}\ }\textbf {\bibinfo {volume} {G33}},\ \bibinfo {pages}
  {R111} (\bibinfo {year} {2007})},\ \Eprint
  {http://arxiv.org/abs/nucl-th/0612100} {arXiv:nucl-th/0612100 [nucl-th]}
  \BibitemShut {NoStop}%
\bibitem [{\citenamefont {Moore}\ \emph {et~al.}(2015)\citenamefont {Moore},
  \citenamefont {Cole},\ and\ \citenamefont {Berry}}]{Moore:2014lga}%
  \BibitemOpen
  \bibfield  {author} {\bibinfo {author} {\bibfnamefont {C.~J.}\ \bibnamefont
  {Moore}}, \bibinfo {author} {\bibfnamefont {R.~H.}\ \bibnamefont {Cole}}, \
  and\ \bibinfo {author} {\bibfnamefont {C.~P.~L.}\ \bibnamefont {Berry}},\
  }\href {\doibase 10.1088/0264-9381/32/1/015014} {\bibfield  {journal}
  {\bibinfo  {journal} {Class. Quant. Grav.}\ }\textbf {\bibinfo {volume}
  {32}},\ \bibinfo {pages} {015014} (\bibinfo {year} {2015})},\ \Eprint
  {http://arxiv.org/abs/1408.0740} {arXiv:1408.0740 [gr-qc]} \BibitemShut
  {NoStop}%
\bibitem [{\citenamefont {Abbott}\ \emph
  {et~al.}(2019{\natexlab{a}})\citenamefont {Abbott} \emph
  {et~al.}}]{Authors:2019ztc}%
  \BibitemOpen
  \bibfield  {author} {\bibinfo {author} {\bibfnamefont {B.~P.}\ \bibnamefont
  {Abbott}} \emph {et~al.} (\bibinfo {collaboration} {LIGO Scientific,
  Virgo}),\ }\href {\doibase 10.3847/1538-4357/ab20cb} {\bibfield  {journal}
  {\bibinfo  {journal} {Astrophys. J.}\ }\textbf {\bibinfo {volume} {879}},\
  \bibinfo {pages} {10} (\bibinfo {year} {2019}{\natexlab{a}})},\ \Eprint
  {http://arxiv.org/abs/1902.08507} {arXiv:1902.08507 [astro-ph.HE]}
  \BibitemShut {NoStop}%
\bibitem [{\citenamefont {Abbott}\ \emph
  {et~al.}(2019{\natexlab{b}})\citenamefont {Abbott} \emph
  {et~al.}}]{Abbott:2019bed}%
  \BibitemOpen
  \bibfield  {author} {\bibinfo {author} {\bibfnamefont {B.~P.}\ \bibnamefont
  {Abbott}} \emph {et~al.} (\bibinfo {collaboration} {LIGO Scientific,
  Virgo}),\ }\href@noop {} {\  (\bibinfo {year} {2019}{\natexlab{b}})},\
  \Eprint {http://arxiv.org/abs/1902.08442} {arXiv:1902.08442 [gr-qc]}
  \BibitemShut {NoStop}%
\bibitem [{\citenamefont {Folacci}\ and\ \citenamefont {Ould
  El~Hadj}(2019)}]{Folacci:2019cmc}%
  \BibitemOpen
  \bibfield  {author} {\bibinfo {author} {\bibfnamefont {A.}~\bibnamefont
  {Folacci}}\ and\ \bibinfo {author} {\bibfnamefont {M.}~\bibnamefont {Ould
  El~Hadj}},\ }\href@noop {} {\  (\bibinfo {year} {2019})},\ \Eprint
  {http://arxiv.org/abs/1901.03965} {arXiv:1901.03965 [gr-qc]} \BibitemShut
  {NoStop}%
\bibitem [{\citenamefont {Bontz}\ and\ \citenamefont
  {Haugan}(1981)}]{Bontz:1981}%
  \BibitemOpen
  \bibfield  {author} {\bibinfo {author} {\bibfnamefont {R.~J.}\ \bibnamefont
  {Bontz}}\ and\ \bibinfo {author} {\bibfnamefont {M.~P.}\ \bibnamefont
  {Haugan}},\ }\href@noop {} {\bibfield  {journal} {\bibinfo  {journal}
  {Astrophysics and Space Science}\ }\textbf {\bibinfo {volume} {78}},\
  \bibinfo {pages} {199} (\bibinfo {year} {1981})}\BibitemShut {NoStop}%
\bibitem [{\citenamefont {Berry}(2005)}]{Berry:2005:tsunami}%
  \BibitemOpen
  \bibfield  {author} {\bibinfo {author} {\bibfnamefont {M.~V.}\ \bibnamefont
  {Berry}},\ }\href@noop {} {\bibfield  {journal} {\bibinfo  {journal} {New
  Journal of Physics}\ }\textbf {\bibinfo {volume} {7}},\ \bibinfo {pages}
  {129} (\bibinfo {year} {2005})}\BibitemShut {NoStop}%
\bibitem [{\citenamefont {Berry}(2007)}]{Berry:2007:focused}%
  \BibitemOpen
  \bibfield  {author} {\bibinfo {author} {\bibfnamefont {M.}~\bibnamefont
  {Berry}},\ }in\ \href@noop {} {\emph {\bibinfo {booktitle} {Proceedings of
  the Royal Society of London A: Mathematical, Physical and Engineering
  Sciences}}},\ Vol.\ \bibinfo {volume} {463}\ (\bibinfo {organization} {The
  Royal Society},\ \bibinfo {address} {London, UK},\ \bibinfo {year} {2007})\
  pp.\ \bibinfo {pages} {3055--3071}\BibitemShut {NoStop}%
\bibitem [{\citenamefont {Isaacson}(1968)}]{Isaacson:1968zza}%
  \BibitemOpen
  \bibfield  {author} {\bibinfo {author} {\bibfnamefont {R.~A.}\ \bibnamefont
  {Isaacson}},\ }\href {\doibase 10.1103/PhysRev.166.1272} {\bibfield
  {journal} {\bibinfo  {journal} {Phys. Rev.}\ }\textbf {\bibinfo {volume}
  {166}},\ \bibinfo {pages} {1272} (\bibinfo {year} {1968})}\BibitemShut
  {NoStop}%
\bibitem [{\citenamefont {Brill}\ and\ \citenamefont
  {Hartle}(1964)}]{Brill:1964zz}%
  \BibitemOpen
  \bibfield  {author} {\bibinfo {author} {\bibfnamefont {D.~R.}\ \bibnamefont
  {Brill}}\ and\ \bibinfo {author} {\bibfnamefont {J.~B.}\ \bibnamefont
  {Hartle}},\ }\href {\doibase 10.1103/PhysRev.135.B271} {\bibfield  {journal}
  {\bibinfo  {journal} {Phys. Rev.}\ }\textbf {\bibinfo {volume} {135}},\
  \bibinfo {pages} {B271} (\bibinfo {year} {1964})}\BibitemShut {NoStop}%
\end{thebibliography}%
\bibliographystyle{apsrev4-1}

\end{document}